\documentclass{aastex61}
\usepackage{amsmath}
\usepackage{color}

\shorttitle{AQ Col (EC~05217--3914)}
\shortauthors{Otani et al.}
\graphicspath{{./}{figures/}}

\accepted{Nov 15, 2021}
\submitjournal{ApJ}

\begin{document}
\title{Anomalous Orbital Characteristics of the AQ~Col (EC~05217--3914) System} 

\correspondingauthor{T. Otani}
\email{otanit@erau.edu}

\author[0000-0002-0786-7307]{T. Otani}
\affiliation{Department of Physical Sciences and SARA, Embry-Riddle Aeronautical University, 1 Aerospace Blvd, Daytona Beach, FL 32114, USA}

\author{A. E. Lynas-Gray}
\affiliation{Department of Physics and Astronomy, University College London, Gower Street, London WC1E 6BT, United Kingdom}
\affiliation{Department of Physics, University of Oxford, Keble Road, Oxford OX1 3RH, United Kingdom}
\affiliation{Department of Physics and Astronomy, University of the Western Cape, Bellville 7535, South Africa}

\author{D. Kilkenny}
\affiliation{Department of Physics and Astronomy, University of the Western Cape, Bellville 7535, South Africa}

\author{C. Koen}
\affiliation{Department of Statistics, University of the Western Cape, Bellville 7535, South Africa}

\author{T. von Hippel}
\affiliation{Department of Physical Sciences and SARA, Embry-Riddle Aeronautical University, 1 Aerospace Blvd, Daytona Beach, FL 32114, USA}

\author{M. Uzundag}
\affil{Instituto de F\'isica y Astronom\'ia, Universidad de Valpara\'iso, Gran Breta\~na 1111, Playa Ancha, Valpara\'iso, 2360102, Chile}
\affil{European Southern Observatory, Alonso de Cordova 3107, Santiago, Chile}

\author{M. V\v{u}ckovi\'{c}}
\affil{Instituto de F\'isica y Astronom\'ia, Universidad de Valpara\'iso, Gran Breta\~na 1111, Playa Ancha, Valpara\'iso, 2360102, Chile}

\author{C. M. Pennock}
\affil{Lennard-Jones Laboratories, Keele University, ST5 5BG, United Kingdom}

\author{R. Silvotti}
\affil{INAF-Osservatorio Astrofisico di Torino, strada dell'Osservatorio 20, 10025 Pino Torinese, Italy}

\begin{abstract}

AQ~Col (EC~05217-3914) is one of the first detected pulsating subdwarf B (sdB) stars and has been considered to be a single star.  
Photometric monitoring of AQ~Col reveals a 
pulsation timing variation with a period of 486~days, 
interpreted as time-delay due to reflex motion in a wide-binary formed
with an unseen companion with expected mass larger than 1.05 $M_\odot$. The optical spectra and color-Magnitude diagram of the system suggested that the companion is not a main sequence star but a white dwarf or neutron star. The pulsation timing variation also shows that the system has an eccentricity of 0.424, which is much larger than any known sdB long period binary system. That might be due to the existence of another short period companion to the sdB star. Two optical spectra obtained on
1996 December $5^{\rm th}$ show a radial velocity change of 49.1~km/s 
in 46.1 minutes, which suggests the hot subdwarf in the wide-binary is
itself a close-binary formed with another unseen white dwarf or neutron star companion; if further observations show this interpretation
to be correct, AQ~Col is an interesting triple system worthy of further
study.
                   
\end{abstract}

\section{Introduction} \label{sec:intro}
Subdwarf B (sdB) stars are core helium burning objects, found in both the disk and halo of our Galaxy \citep{Saffer et al.1994}. EC 14026-2647 (= V361 Hya) was the first pulsating sdB star to be discovered \citep{Kilkenny et al.1997}, possibly because, as suggested by \citet{Ostensen et al.2010}, only about 10$\%$ of these pressure mode (p-mode) sdB stars have pulsations detectable from the ground.  The observed properties of sdB stars place them on the extreme horizontal branch (EHB). Their effective temperatures range from 22,000 to 40,000 $K$ and surface gravities range from 5.0 $\le$ log $g$ $\le$6.2 (in cgs units).  Their masses are narrowly confined to about 0.5$M_{\sun}$ \citep{Heber2009}.  Subdwarf B stars have experienced mass-loss at, or just before the end of the red giant branch phase \citep{Bonanno et al.2003}, in which the hydrogen envelope is lost, leaving a helium core with a very thin inert hydrogen-rich envelope. The loss of the hydrogen envelope prevents the stars from entering the Red Clump phase and subsequently ascending the Asymptotic Giant Branch; instead, and they settle on the EHB, spending about $10^8$ years as sdB stars. Upon helium depletion in the core they become subdwarf O (sdO) stars burning helium in a shell surrounding a C/O core and, eventually, DAO white dwarfs \citep{Dorman et al.1993, Bergeron et al.1994}. 
      
Plausible sdB formation models via binary evolution were constructed by \citet{Han et al.2002, Han et al.2003} and more than 50\% of hot subdwarfs are understood to be members of binary systems (see e.g. \citet{Silvotti et al.2021} and references therein). However, some fraction of sdB stars may not be in binaries \citep{Heber2009, Fontaine et al.2012}.  If true, this would require another formation channel, perhaps the single star evolution scenario proposed by both \citet{Dorman et al.1993} and \citet {DCruz1996}. In another scenario, the merger of a helium white dwarf with a low-mass hydrogen-burning star was proposed as a way of forming single sdB stars \citep{Clausen et al.2011}. \citet{Wu2018, Wu2020} recently suggested the possibility that sdB + neutron stars (NS) binary systems should contribute 0.3\% to 0.5\% of the total sdB binaries.  To distinguish between these evolutionary scenarios, orbital information on sdB star binaries is essential. 
      
The existence of sdB pulsators (sdBV, where the "V" stands for variable), adopting the classification scheme \citet{Kilkenny et al.2010} propose, was predicted by \citet{Charpinet et al.1996}.  Independently, \citet{Kilkenny et al.1997} discovered the first short period sdBV star, \objectname{EC 14026-2647}.  These stars are p-mode pulsators, where pulsations are driven by internal pressure fluctuations \citep{Charpinet et al.2000}.  The first long period sdBV star, \objectname{PG 1716+426}, is a g-mode pulsator \citep{Green et al.2003}, in which gravity provides the restoring force.  Some sdB stars have been discovered to exhibit both p-mode and g-mode pulsations.  These objects are called hybrid pulsators \citep{Schuh et al.2006, Oreiro et al.2004}. 
      
Although amplitudes often vary (e.g. \citealp{Kilkenny2010}), the pulsation periods of sdBV stars are usually stable \citep{Ostensen et al.2001}, and therefore they are good chronometers.  As is well known, a star's position in space may show cyclic variations due to the gravitational perturbations of a companion.  From an observer's point of view, the light from the pulsating star is periodically delayed when it is on the far side of its orbit and advanced on the near side. The orbital solution of the binary system can be obtained from changes in the pulsation arrival times, and this is referred to as the pulsation timing method. This technique has long been used in the binary star community to search for additional components, orbital period changes, mass loss, etc.
      
Several candidate planets and companions to sdB host stars have been detected by this method. \citet{Silvotti et al.2007, Silvotti et al.2018} detected a planet candidate around the sdB star \objectname{V391 Peg} in this way.  \citet{Lutz2011} detected companions to the sdB stars \objectname{HS 0444+0458} and \objectname{HS 0702+6043} which appear to be a brown dwarf and an exoplanet, respectively. However, \citet{Mackebrandt et al.2020} did not confirm these tentative determinations.  \citet{Mullally et al.2008} used this Observed minus Calculated (O-C) method to search for possible planets around DAV white dwarfs. Among the 15 white dwarf stars they surveyed, GD 66 exhibited O-C variations consistent with a 2 $M_{J}$ planet in a 4.5-year orbit which, however, was not confirmed by \citet{Dalessio2013}. \citet{Zong et al.2016, Zong et al.2018} recently pointed out that the pulsation frequencies of sdB stars and white dwarfs show variations and that requires us to be more cautious in using the pulsation timing method to search for small mass objects like planets. However, this method is still an efficient tool to obtain an orbital solution of a binary system. The presence of M-dwarf or white dwarf companion to the sdBV star CS1246, suggested by O-C variations, was confirmed by RV measurements \citep{Barlow et al.2011a, Barlow et al.2011b}.  Similarly, \citet{Otani et al.2018} used this method to obtain an orbital solution of the previously known sdB and main sequence binary, EC 20117-4014. It is perhaps worth noting here that \citet{Bours et al.2016} present O-C diagrams for more than 50 white dwarf binaries and suggest that the period variations might be due to magnetic effects (such as the Applegate mechanism \citep{1992ApJ...385..621A}) in the cool companions.  However, where there are substantial baselines, these binaries do not show repeatable cyclic variations and the timescales of the effects are much longer than those described here.

\objectname[AQ~Col]{AQ~Col (EC 05217-3914)} is an sdB star originally found in the Edinburgh-Cape (EC) Blue Object Survey \citep{Stobie et al.1997, Kilkenny et al.2016}. The spectra obtained by \citet{Koen et al.1999} did not suggest that \objectname[AQ~Col]{AQ~Col} has a companion. The apparent magnitude of the star is V = 15.55 \citep{Koen et al.1999}. Three pulsation periods of 218 s, 216 s, and 213 s were detected, making \objectname[AQ~Col]{AQ~Col} one of the first members of the class of short period sdBV stars.  Further high speed photometry of \objectname[AQ~Col]{AQ~Col} was obtained and an asteroseismic analysis was performed by \citet{Billeres2005}, who estimated the effective temperature ($T_{eff}$ = 32,000 K), gravity (log $g$ = 5.730), and mass ($M_{*}$ = 0.49 $M_{\sun}$). They also found two additional pulsation periods at 208 s and 129 s.

Although \objectname[AQ~Col]{AQ~Col} has been believed to be a single sdB star for decades, our pulsation timing analysis using the three largest amplitude pulsations show periodic variations, which indicate that AQ~Col has a companion with a long orbital period. So far, we have photometric data spread over nearly 25 years for AQ~Col. In this paper, we present the results of our pulsation timing analysis, and the extracted orbital information for the system. We also analyzed the spectroscopic data of \citet{Koen et al.1999} and an additional spectrum obtained in 2020.  The radial velocity differences obtained from the spectroscopic data indicate that AQ~Col may have another companion with a short orbital period, which we also discuss. 
      
Section~\ref{sec:obser} outlines the facilities and instrumentation used to obtain the data needed for the analysis and our reduction procedures.  Section~\ref{sec:OCmethod} shows the pulsation timing method that we used. Section~\ref{sec:results} presents our results derived from the observed pulsation peaks in the frequency spectrum of \objectname[AQ~Col]{AQ~Col} and spectroscopy data.   Our conclusions and suggested additional work are summarised in Section~\ref{sec:conc}.



\section{Observations and Data Reduction} \label{sec:obser}
\subsection{Photometry Data}

The observation log is presented in Table~\ref{tab:table2}.  For observations made at the Chilean site of the Southeastern Association for Research in Astronomy (SARA-CT), no filter was used since the target was faint for the 0.6m SARA-CT telescope\footnote{SARA-CT 0.6 m telescope is owned and operated by the Southeastern Association for Research in Astronomy (saraobservatory.org).} \citep{Keel et al.2017}. To reduce read-out noise, 3x3 on-chip binning was used for the images.  The exposure time was 30 s until Feb 2nd, 2018 but reduced to 15 s after Feb 26, 2018. The pulsations frequencies and amplitudes are expected to be wavelength-dependent (see \citet{Koen1998}).  However, separate analyses were performed using only the SAAO data, and both SAAO and SARA-CT data to compare the results.  The pulsation timing analysis results with and without the SARA-CT data were the same within the uncertainty. Therefore, the results presented in this paper include the data taken with SARA-CT.  

For the calibration and reduction of SARA-CT data, standard procedures were performed using AstroImageJ \citep{Collins et al.2017}\footnote{https://www.astro.louisville.edu/software/astroimagej}.  All flat fields were exposed in a twilight sky.  For each night's data, the aperture that gave the best signal-to-noise ratio (S/N) was chosen and sky annuli were used to subtract the sky background.  These values were then divided by similarly extracted intensity values of non-variable comparison stars.   

The early data in Table~\ref{tab:table2} were obtained using a mixture of
the SAAO 1m telescope with a photomultiplier-based photometer, or the SAAO 0.75m and 1m telescopes with the UCT CCD photometer
(see \citet{Koen et al.1999}). All later observations were made with the
SAAO 1m telescope with different CCD photometers - between 2006 and mid-2012 using the UCT CCD;
between mid-2012 and 2019 using the SAAO STE3 CCD; and post-2019,
the STE4 CCD -- the undesirable change of CCDs being 
forced by the demise of the earlier instruments. The photomultiplier
data were reduced by subtracting a cubic-spline fit to occasional
sky measurements from each star measurement and then correcting for
atmospheric extinction (see \citet{O'Donoghue et al.1997} for details). The
CCD data were reduced using an automated version of the DoPhot software \citep{Schechter et al.1993}. A detailed 
description of both the UCT CCD and the CCD reduction process is 
given in \citet{O'Donoghue et al.1996} and information on the STE3 and STE4 photometers can be found on the SAAO web page\footnote{www.saao.ac.za/astronomers/ste3-ste4/}. 

For each night, the raw light curves were then normalized to the mean magnitude for that night. A quadratic curve was used to remove mild curvature in the light curves caused by differential extinction between the target and comparison stars.  All times were corrected to Barycentric Julian Date (BJD) in Barycentric Dynamical Time \citep{Eastman et al.2010}.  

The normalized light curves were then analyzed using Period04 \citep{Lenz2004}.  To improve detection and characterization of the pulsation frequencies and amplitudes, the data were pre-whitened \citep{Blackman1958}.  After one or more frequencies were identified in the amplitude spectrum, they were removed from each light curve by subtracting the corresponding least-squares fitted sine curve \citep{Sullivan et al.2008}.  This analysis was performed for each of the runs listed in Table~\ref{tab:table2}. As described in
Section 4, we only use peaks above 4$\sigma$ for our analysis to find an orbital solution.




\startlongtable
\begin{deluxetable}{cccc}
\tablecaption{Observation Log for AQ~Col \label{tab:table2}. Only the observation dates in which the data show any pulsation amplitude $\geq$ 2$\sigma$ above the noise level are shown.  The 1996 and 1998 data were previously published in \citet{Koen et al.1999}.}
\tablehead{\colhead{Date} & \colhead{Mean Time} & \colhead{Observation Length} & \colhead{Observatory}\\ 
\colhead{} & \colhead{(BJD$-$2453500)} & \colhead{(min)} & \colhead{} } 
\startdata
1996 Jan 28  & $-$3388.60  & 252.2           & SAAO        \\
1996 Feb 17  & $-$3368.63  & 217.2           & SAAO        \\
1996 Dec 5  & $-$3076.56  & 460.4           & SAAO        \\
1996 Dec 9  & $-$3072.57  & 422.5           & SAAO        \\
1998 Jan 27  & $-$2658.59  & 304.4           & SAAO        \\
1998 Jan 28  & $-$2657.59  & 302.2           & SAAO        \\
1998 Feb 1   & $-$2653.61  & 329.1           & SAAO        \\
2006 Dec 19 & 589.38    & 264.1           & SAAO        \\
2007 Feb 18  & 650.33    & 164.5           & SAAO        \\
2010 Oct 15 & 1984.79   & 242.4           & SARA-S      \\
2010 Dec 9  & 2039.70   & 389.2           & SARA-S      \\
2010 Dec 10 & 2040.69   & 420.0           & SARA-S      \\
2015 Nov 7  & 3833.74   & 319.4           & SARA-S      \\
2015 Dec 8  & 3864.69   & 438.9           & SARA-S      \\
2016 Feb 7   & 3925.70   & 151.5           & SARA-S      \\
2017 Oct 17 & 4544.56   & 188.0           & SAAO        \\
2017 Oct 20 & 4546.52   & 86.8            & SAAO        \\
2017 Oct 21 & 4547.55   & 192.4           & SAAO        \\
2018 Jan 15  & 4633.72   & 157.0           & SARA-S      \\
2018 Jan 19  & 4637.64   & 210.2           & SARA-S      \\
2018 Feb 2   & 4651.61   & 178.1           & SARA-S      \\
2018 Feb 26  & 4675.61   & 159.9           & SARA-S      \\
2018 Mar 14  & 4692.29   & 134.5           & SAAO        \\
2018 Mar 16  & 4694.28   & 116.0           & SAAO        \\
2018 Mar 17  & 4695.29   & 138.1           & SAAO        \\
2018 Mar 19  & 4697.28   & 114.0           & SAAO        \\
2018 Nov 8  & 4930.55   & 152.4           & SAAO        \\
2018 Nov 9  & 4931.51   & 159.2           & SAAO        \\
2018 Nov 10 & 4932.55   & 128.6           & SAAO        \\
2018 Nov 12 & 4934.53   & 223.2           & SAAO        \\
2019 Jan 16  & 4999.70   & 316.6           & SARA-S      \\
2019 Feb 6   & 5021.34   & 183.7           & SAAO        \\
2019 Mar 6   & 5049.31   & 165.9           & SAAO        \\
2019 Mar 12  & 5055.30   & 148.4           & SAAO        \\
2019 Oct 29 & 5055.30   & 155.2           & SAAO        \\
2019 Nov 20 & 5308.51   & 136.5           & SAAO        \\
2019 Nov 22 & 5310.46   & 162.5           & SAAO        \\
2019 Nov 23 & 5311.50   & 234.5           & SAAO        \\
2019 Nov 24 & 5312.48   & 170.9           & SAAO        \\
2019 Nov 26 & 5314.51   & 213.9           & SAAO        \\
2020 Mar 13  & 5422.30   & 119.5           & SAAO        \\
2020 Mar 21  & 5430.28   & 124.3           & SAAO     
\\
\enddata
\end{deluxetable}

\subsection{Spectroscopic Data}

Spectra were obtained with the South African Astronomical Observatory
(SAAO) 1.9-m telescope as part of the Edinburgh-Cape (EC) Survey as
\citet{Stobie et al.1997} describe.  As was standard practice for
early EC~Survey spectroscopy, a Reticon Spectrograph
\citep{1982SPIE..331..368J} was used with Grating-6 and a 250$\mu$ slit,
corresponding to 1.8~arcsec and giving an effective resolution
(full-width at half-maximum) of FWHM $\simeq 3.5$\AA.  As was customary,
a 100-s Cu/Ar arc spectrum was obtained before and after each sequence:
star and sky spectra were obtained using separated detectors with an exposure 
time of 1200-s, and their role reversed for a second 1200-s exposure,
with a third arc-spectrum obtained between the two.  A single 
wavelength-calibrated AQ~Col spectrum corrected for 
sky-background was thereby obtained,
having a useful wavelength range of $3600 < \lambda < 5200\,$\AA.
Spectra available for analysis are listed in 
Table \ref{tab:number5}, the sequence described above being used to
secure all three SAAO spectra.

An additional spectrum ($3700 < \lambda < 7200\,$\AA) was observed with
the Southern Astrophysical Research (SOAR) Telescope using the Goodman
Spectrograph \citep{2004SPIE.5492..331C}; in this case
the effective resolution was FWHM $\simeq 2.0$\AA.  Barycentric radial
velocities included in Table \ref{tab:number5}
were obtained by cross-correlating against a synthetic spectrum for
the \citet{Koen et al.1999} atmospheric parameters
$T_{\rm eff} = 31000$K, $\log g = 5.7$, and 
$\log (N({\rm He})/N({\rm H})) = -5.0$, taken from the 
\citet{2014ASPC..481...95N} non-LTE grid; the
$\log (N({\rm He})/N({\rm H}))$ choice being based on the absence
of He~I lines.  Barycentric corrections were obtained
following \citet{2014PASP..126..838W}.

\begin{deluxetable}{cccc}[ht!]
\caption{Spectra Available for Analysis}
\label{tab:number5}
\tablehead{Observation Date & Telescope  &  HJD -- 2440000 & Radial Velocity  \\
                            &            & (mid-exposure)  &     (km/s)        }
\startdata
1989 December $21$ & SAAO 1.9-m & 07881.52318 & $-89.2$ \\
1996 December $05$ & SAAO 1.9-m & 10422.53204 & $+169.4$ \\
1996 December $05$ & SAAO 1.9-m & 10422.56408 & $+120.3$ \\
2020 February $28$ & SOAR 4.1-m & 18907.55403 & $+199.8$ \\
\enddata
\end{deluxetable}


\section{pulsation timing  method} \label{sec:OCmethod}

Stable light curve variations shown by pulsations or eclipses can act as accurate clocks, and monitoring those timings allows us to search for phenomena such as the existence of planets and companions, stellar evolution, or apsidal motion of the binary system. This is one of the classic techniques in astronomy, and this principle was used by R{\o}mer in the late 17th century to observe apparent periodic changes in Jupiter's Galilean moons and thus to estimate the speed of light.   

To monitor timing variations in the light curve oscillations, the Observed minus Calculated (O-C) method is the most common method. Computing the time difference between an observed event and that calculated from an ephemeris allows one to determine an accurate period of the periodic event and to search for cyclic and secular variations. As is standard practice, the observed times of the light curve maxima are used to form the ephemeris and then the (O-C) values are plotted as a function of time.  Good reviews of this method can be found in \citet{Paparo et al.1988}, \citet {Sterken2005}, and \citet{Winget2008}. The traditional O-C method uses the light curve maxima to obtain the timing variations, however, monitoring pulsation phase changes will also let one calculate the same timing variations \citep{Murphy et al.2014}.  This approach is called the Phase Modulation (PM) method and is particularly suited for analysis of multimode pulsators; it also allows us to obtain a better quality for the timing variations because all the data are used. Since AQ~Col is a multimode pulsator, the PM method was used in this paper. However, the basic concept (obtaining the timing variations) and how to interpret the variations to the astronomical phenomenon are the same for both the O-C method and the PM method. 

The pulsation timing variations, $\tau$, can be expressed as a quadratic;

\begin{equation}
\tau = \frac{1}{2}P\dot{P}E^2 + {\Delta}P{\,}E + \Delta E_0,
\label{eq:evolution}
\end{equation}

\noindent where $E$ is the integer counter of completed cycles after the first observation, $P$ is the initially estimated period of pulsation, $\Delta E_{0}$ is the  difference between the observed and calculated reference epochs, $\Delta P$ is the difference between the actual period and the estimated period, and $\dot{P} = dP/dt$ (see \citet{Sterken2005} and \citet{Winget2008} for details). The pulsation timing variations (this concept also works for the other timing methods such as the eclipse timing method) will be constant if no pulsation period changes are occurring and the assumed pulsation period is correct.  If the calculated period is constant but incorrect, $\tau$ will be linear with a positive or negative slope.  If the period is changing linearly with time (e.g. due to the star evolving or magnetic braking), $\tau$ will exhibit a quadratic form. The precision of this technique, when applied to observations spanning several years, has allowed empirical measurement of the cooling rates of white dwarf stars and the evolution of sdB stars \citep{Kepler1991, Silvotti et al.2007,Winget2008,Costa2008, Lutz2011,Barlow et al.2011c,Otani et al.2018, Kepler2021} and other evolved stars \citep {Kilkenny et al.2005}. 

 If the pulsation timing variations, $\tau$, show periodicities, they are most likely caused by the beating of two closely spaced pulsation frequencies or reflex motion due to an unseen companion.  The beating of two closely spaced frequencies, which may not be resolved in the power spectrum, causes not only sinusoidal variability in the pulsation timing but also sinusoidal variability in pulsation $amplitudes$ with a phase difference of 90 degrees \citep{Kepler1983, Lutz2011}.  If the pulsation timing variations are caused by reflex motion, the orbital solutions of the binary (or the planetary) system can be obtained from the variations. Searching for orbital solutions using this timing method has been discussed for a century since \citet{Woltjer1922} first determined the elliptic orbit of a binary star. \citet{Irwin1952,Irwin1959} showed that an apparent variation of an 
eclipsing binary (EB) period could be caused by a changing light-travel-time due to the reflex motion of its centre-of-mass caused by the orbital motion of a generally more distant third star in the system.  \textbf{Recently, \citet{Kurtz et al.2015} obtained the orbital solution of the unseen companion to an eclipsing binary star KIC 8569819 using this technique to verify the frequency modulation (FM) method. The FM method is similar to the PM method but it uses frequency modulation to find the orbital solution of the unseen companion.}

 The distance $z$ between the object of interest and the binary system's center of gravity is generally described as

\begin{equation}
z =  \frac{a_1 \sin{i} (1-e^2) \sin{(f+\varpi)}}{1+e\cos{f}},
\label{eq:z}
\end{equation}
where $a_1$ is the length of the pulsating star orbit semi-major axis, $e$ is the orbital
eccentricity, $f$ is the true anomaly, and $\varpi$ is the argument of periapsis (e.g. \citet{Smart1977}). The pulsation timing variations $(\tau = z/c)$
are largest when absolute values of $z$ are also maxima, where $c$ is the speed of light. 

\textbf{The true anomaly changes over time. When e \(\ll\) 1 (close to a circular orbit), f changes linearly over time, and is described by} $f = 2\pi\nu_{orb}(t-t_0)$, where, $\nu_{orb}$ is the orbital frequency of the sdB star, $t$ is the time, and $t_0$ is the time when the pulsating star passed the argument of periapsis. Therefore, the pulsation timing variation as a function of time can be shown to be

 \begin{equation}
\tau(t) =  \frac{a_1 \sin{i}}{c}\frac{(1-e^2)\sin(2\pi\nu_{orb}(t-t_0)+\varpi)}{(1+e\cos{2\pi\nu_{orb}(t-t_0)})} \approx \frac{a_1 \sin{i}}{c}\sin(2\pi\nu_{orb}(t-t_0)+\varpi).
\label{eq:binary1}
\end{equation}

When the eccentricity $e$ is not small enough to assume that the orbit is almost circular, the true anomaly $f$ is not constantly changing over time. However, the trigonometric functions of true anomaly $f$ can be described using Bessel functions \citep{Shibahashi2012} as follows:

\begin{equation}
\cos{f} = -e + \frac{2(1-e^2)}{e} \sum_{n=1}^{\infty} J_n(ne) \cos{(n2\pi\nu_{orb}t)}
\label{eq:cosf}
\end{equation}

\begin{equation}
\sin{f} = 2\sqrt{1-e^2} \sum_{n=1}^{\infty} J_n'(ne) \sin{(n2\pi\nu_{orb}t)}
\label{eq:sinf}
\end{equation}
where $J_n(x)$ is the Bessel function of the first kind of integer order $n$, and $J_n'(x)$ = $\frac{dJ_n(x)}{dx}$.

Using these equations, the pulsation timing variation is described as a function of time (see \citet{Murphy et al.2014} for the derivation) as 
\begin{equation}
\tau(t) = \frac{1}{c}a_1\sin{i}[\sum_{n=1}^{\infty}\xi_n(e,\varpi)\sin{(2\pi n\nu_{orb}(t-t_0) + \vartheta_n)}+\tau_0(e,\varpi)]
\label{eq:binary2}
\end{equation}
where

\begin{equation}
a_n(e) = \frac{2\sqrt{1-e^2}}{e}\frac{1}{n} J_n(ne),
\label{eq:an}
\end{equation}

\begin{equation}
b_n(e) = \frac{2}{n} J_n'(ne),
\label{eq:bn}
\end{equation}

\begin{equation}
\xi_n(e,\varpi) = \sqrt{(a_n(e))^2 \cos^2{\varpi} + (b_n(e))^2 \sin^2{\varpi}},
\label{eq:varpi}
\end{equation}

\begin{equation}
\vartheta_n(e,\varpi) = \tan^{-1}{(\frac{b_n(e)}{a_n(e)}\tan{\varpi})}
= \tan^{-1}{(\frac{e}{\sqrt{1-e^2}} \frac{J_n'(ne)}{J_n(ne)}\tan{\varpi})},
\label{eq:vartheta}
\end{equation}
and $\tau_0(e,\varpi)$ is the timing delay at t=0:

\begin{equation}
\tau_0(e,\varpi) = -\sum_{n=1}^{\infty} \xi_n(e,\varpi)\sin{\vartheta_n(e,\varpi)}.
\label{eq:11}
\end{equation}

For the pulsation timing variation of a sdB binary system, both the stellar evolution term, Equation (\ref{eq:evolution}), and the orbital motion term, Equation (\ref{eq:binary2}), should be considered.  Also, the number of cycles, E, in Equation (\ref{eq:evolution}) can be described using time  and pulsation period (E = t/P).  Therefore, the complete expression for $\tau$, is:

\begin{equation}
\tau(t) = \frac{1}{c}a_1\sin{i}\left[\sum_{n=1}^{\infty}\xi_n(e,\varpi)\sin{(2\pi n\nu_{orb}(t-t_0) + \vartheta_n)}+\tau_0(e,\varpi)\right]+ \frac{1}{2} \frac{\dot{P}}{P} t^2 + \frac{\Delta P}{P} t + \Delta E_o.
\label{eq:final}
\end{equation}
In Equations~(\ref{eq:11})~and~(\ref{eq:final}), the summation, $\sum$, will converge absolutely and the values do not significantly change after n = 6 for our data. 

The Fourier Transform of the pulsation timing variation, $\tau$, indicates the orbital frequency $\nu_{orb}$ (frequency at the Fourier Transform peak) and amplitude $ \frac{1}{c}a_1\sin{i}$ (amplitude of the Fourier Transform peak). \citet{Shibahashi2012} and \citet{Murphy et al.2014} suggest that the eccentricity, $e$, can be obtained from the Fourier transform of the $light curve$ of the entire observation runs. In the case of $e \ll 1$, $e$ is  determined from the equations 

\begin{equation}
e \approx \frac{2A_2}{A_1},
\label{eq:e1}
\end{equation}
or 

\begin{equation}
e \approx \frac{4A_3}{3A_2},
\label{eq:e2}
\end{equation}
where $A_1$, $A_2$, and $A_3$ are the amplitudes of the first, second, and third  harmonics. Therefore, $A_1$ is equal to the amplitude of the Fourier transform peak of the light curve. This method to obtain $e$ is suitable particularly for the Kepler targets, that were continuously observed for more than three years. However, for ground-based observations, when the targets are less frequently observed, the noise level of the Fourier Transform of the pulsation timing variations can easily be larger than the amplitude of the harmonics. This was the case for our AQ~Col data, so that $e$ could not be constrained using Equations (\ref{eq:e1}) and (\ref{eq:e2}). Therefore, to constrain the value of $e$, $\varpi$, C~=~$\frac{\dot{P}}{P}$, B~=~$\frac{\Delta P}{P}$, and A~=~$\Delta E_o$, and $t_0$ in Equation~(\ref{eq:final}) the pulsation timing variation data were fitted with Equation  ({\ref{eq:evolution}}) and ({\ref{eq:binary1}}) using the least-squares method. Then the same pulsating timing variation data were fitted again with Equation ({\ref{eq:binary2}}) using the values of $\nu_{orb}$, $e$, $\varpi$, C, B, A, and $t_0$ that are obtained from the previous fit as initial values.

The value $a_1\sin{i}$ obtained from Equation (\ref{eq:binary2}) is used to calculate the mass function \citep{Tauris2006}:

\begin{equation}
f=\frac{(M_{2}\sin{i})^3}{(M_{1}+M_{2})^2}= \frac{1}{2\pi G}K_1^3 P_{orb}(1-e^2)^{3/2} =\frac{(2\pi)^2}{P_{orb}^2G}(a_1\sin{i})^3,
\label{eq:massfunction}
\end{equation}
where $M_1$ and $M_2$ are the masses of the pulsating star and the unseen companion, $P_{orb}$ is the orbital period, G is the gravitational constant, and $K_1 = 2\pi a_1 \sin{i} / P_{orb}\sqrt{1-e^2}$ is the radial velocity (RV) amplitude of the pulsating star. This RV amplitude can be obtained from the time derivative of the position of the star, $v_{rad,1}=-dz/dt$, in which z is written in Equation (\ref{eq:z}). The  derivation of the RV amplitude is clearly described by \citet {Shibahashi2012}, \citet{Shibahashi2015}, and \citet{Murphy2015}. The RV and pulsation timing method are well discussed and used in \citet{Telting+2012 A&A.544.1}, \citet{Telting+2014 A&A.570.129}, \citet{Kurtz et al.2015}, \citet{Murphy+2016 MNRAS.461.4215}, \citet{Nemec+2017 MNRAS.466.1290}, \citet{Lampens+2018 A&A.610.17}, \citet{Murphy+2018 MNRAS.474.4322}, \citet{Derekas+2019 MNRAS.486.2129}, and \citet{Murphy+2021 MNRAS.505.2336}. 

 \section{Results and Discussion} \label{sec:results}    
    \subsection{Pulsation Frequencies Used \label{subsec:pulsation}}
\begin{figure}
\plotone{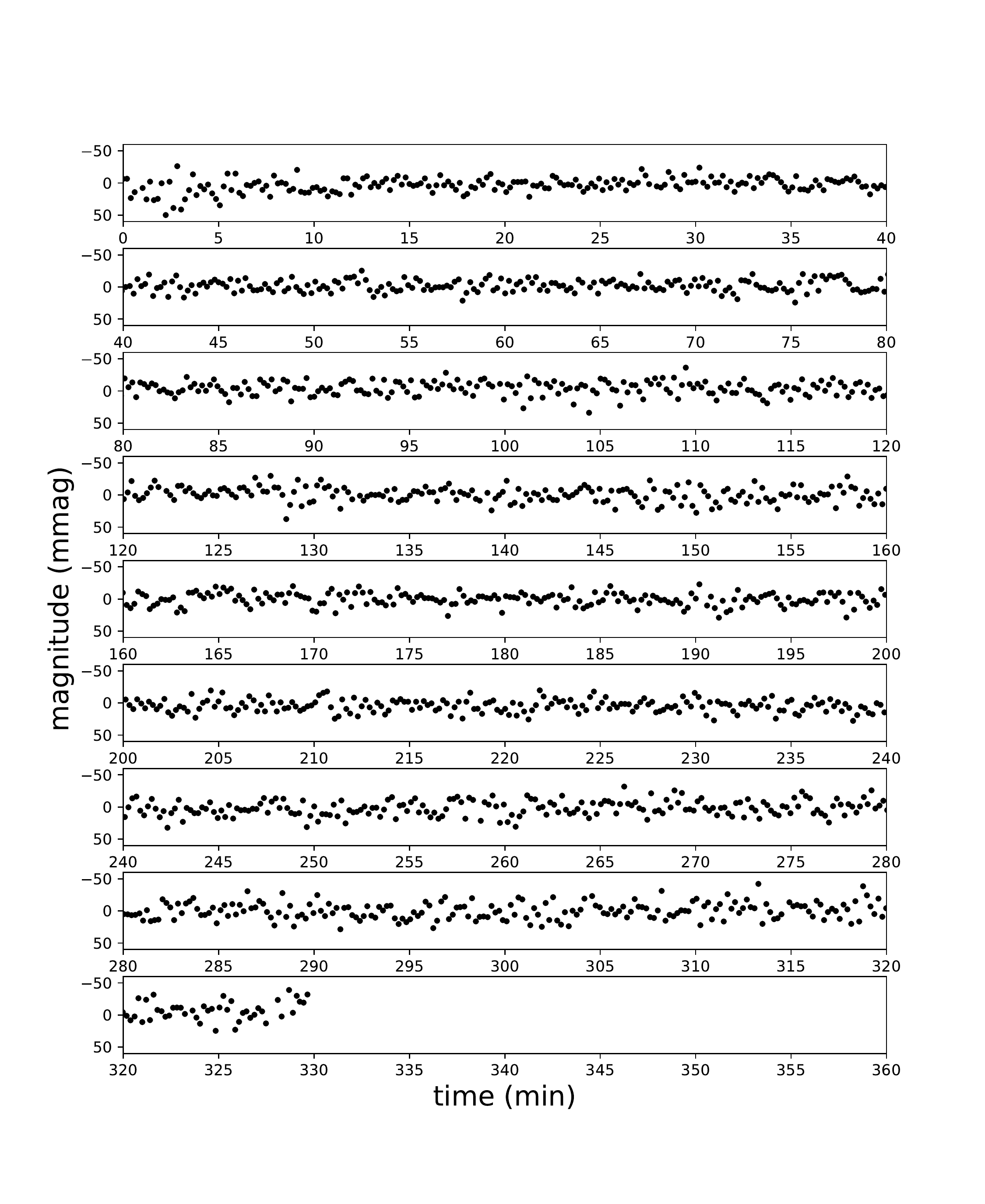} 
\caption{Example light curve for AQ~Col (on Feb 1st, 1998). The lightcurve data of all observation are available as the Data behind the Figure.\label{fig:lc}} 
\end{figure}

\begin{figure}
\plotone{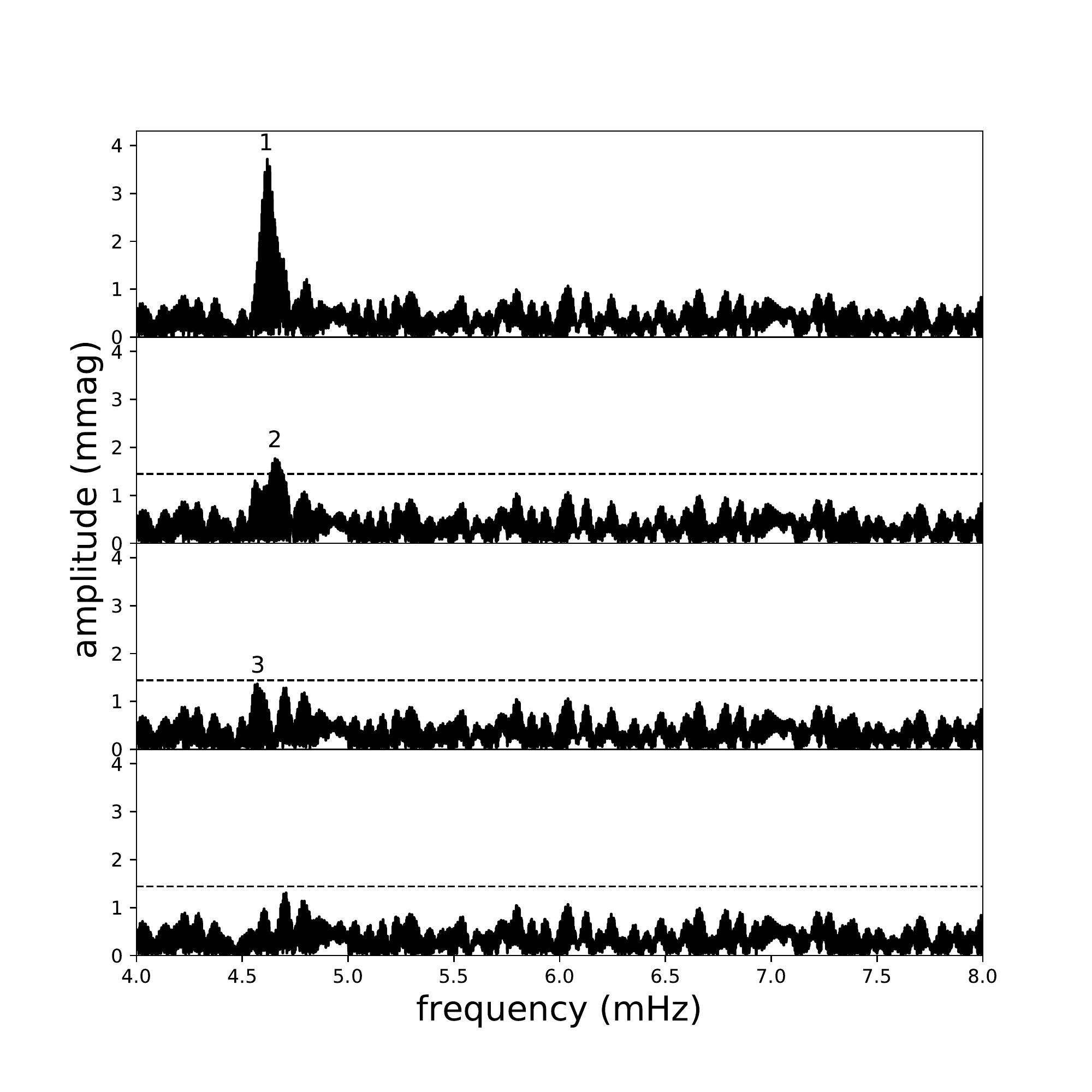} 
\caption{Example Fourier analysis for AQ~Col from one of the best observing runs (combination of Jan 27th, 28th, and Feb 1st, 1998) which indicates the difficulty in extracting the small amplitude pulsations. Three spectral peaks due to pulsation are indicated by numbers. The top panel shows the original Fourier analysis. The lower panels show the successive steps of pre-whitening by sequentially removing the next largest pulsation peaks. In each panel, the horizontal broken lines indicate $4 \sigma$ noise levels; pulsations 1 and 2 are
clearly above these levels while pulsation 3 is below. Noise levels were calculated from the median Fourier amplitude after removing pulsations. Fourier transforms of data from each observing night are shown in the Figure set (5 images), which is available in the online journal. \label{fig:pulsation}} 
\end{figure}

\begin{figure}
\plotone{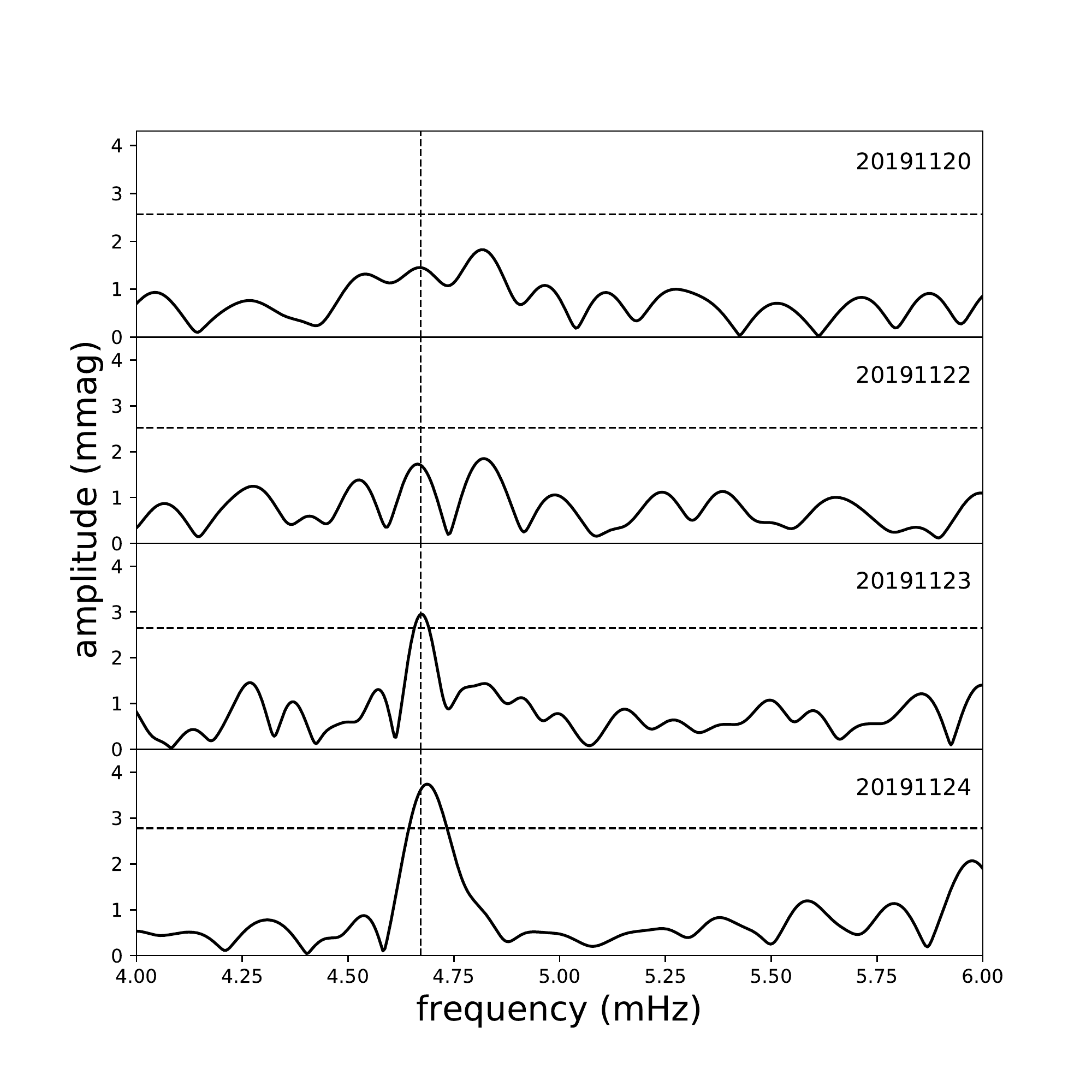} 
\caption{Fourier analysis of four days of observation (Nov 20th - 24th, 2019). The horizontal dashed lines indicate $4\sigma$ noise levels.  The vertical dashed line indicates the F1 (4.67 mHz). The daily amplitude variations of F1 are clearly visible. The F1 pulsation amplitude is below the $4\sigma$ noise level on Nov 20th and 22nd. Noise levels were calculated for each night, therefore the noise level is changing subtly in each panel; note that these vary slightly, reflecting weather changes from night to night and consequent quality variations in resulting light curves. \label{fig:pulsation_daily}}.  
\end{figure}

\begin{figure}
\plotone{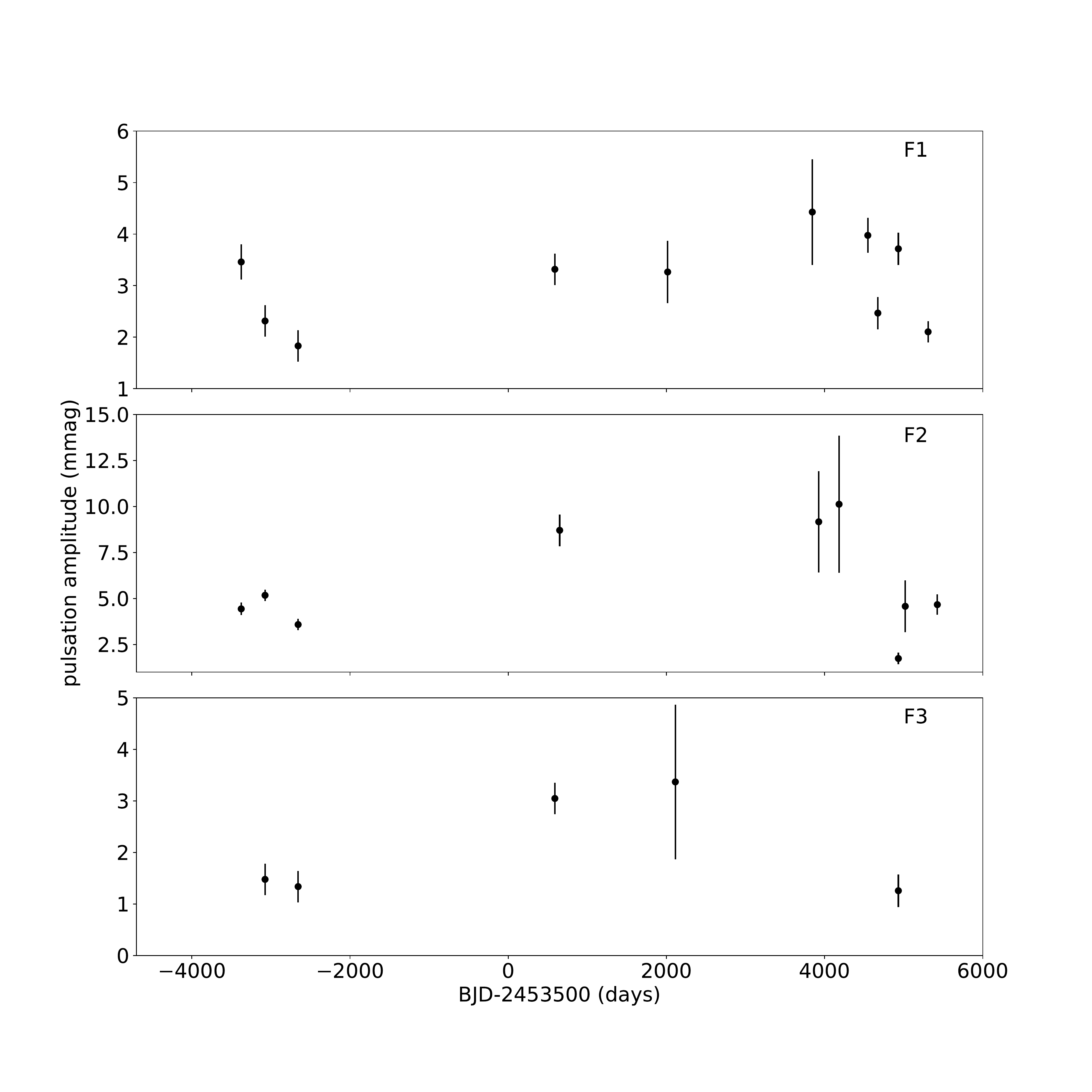} 
\caption{AQ~Col seasonal pulsation amplitude variations of each pulsation mode. This diagram shows that the pulsation amplitude varies each season.  $TESS$ space telescope observed this target in 20 seconds cadence in sector 32 and 33 (BJD-2453500 = 5672 - 5727). However pulsation was not detected above the noise level ($\sim$ 0.3 mmag). \label{fig:pulsation_seasonal}}.  
\end{figure}

\begin{figure}
\plotone{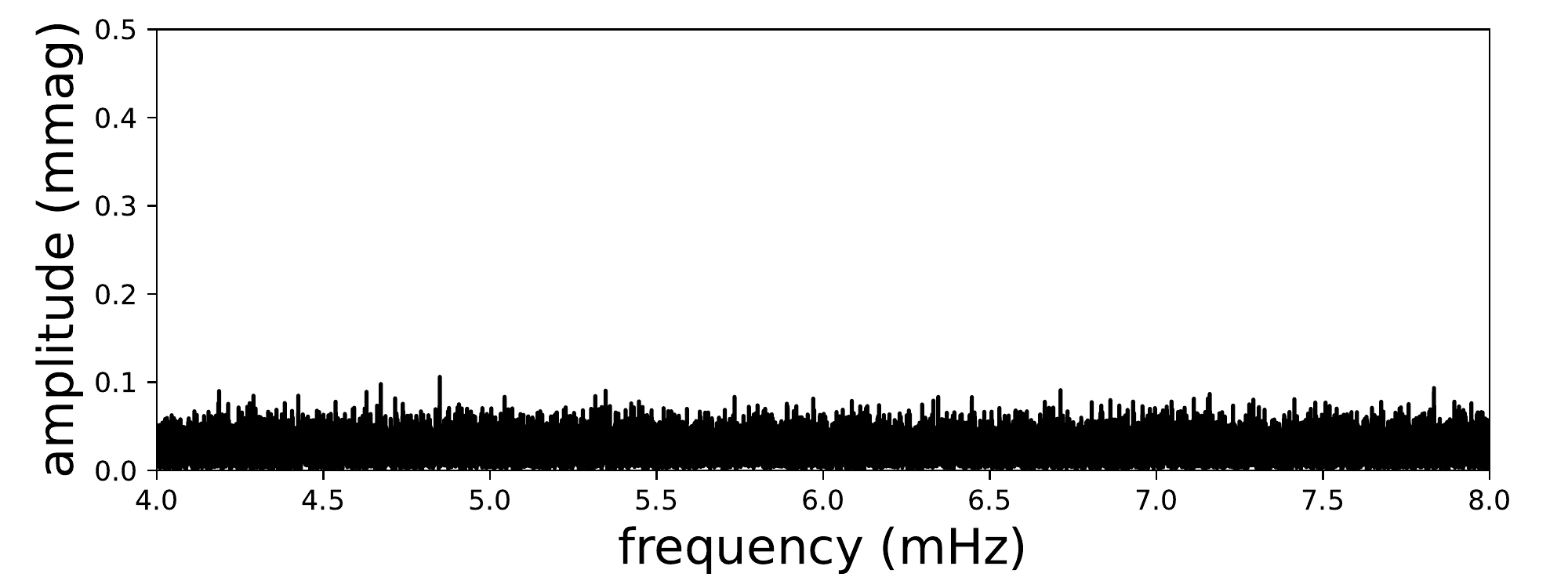} 
\caption{Fourier analysis for AQ~Col from TESS  sectors 32 (Nov 19th, 2020 - Dec 17th, 2020) and 33 (Dec 17th, 2020 - Jan 13th, 2021) observation run. No  peaks due to pulsation are indicated above the noise level (0.03 mmag) \label{fig:pulsation_TESS}} 
\end{figure}

An example light curve for the night of Feb 1st, 1998, and an amplitude spectrum for the combined data from Jan 27th, 28th, and Feb 1st, 1998 are displayed in Figures~\ref{fig:lc} and \ref{fig:pulsation}. Although no pulsation is obvious in the example shown in Figures~\ref{fig:lc}, the amplitude
spectra of the individual data sets consistently recover the same few frequencies but with
clearly variable amplitudes, as described below.  The previously observed pulsation frequency range is 4.3 - 7.7 mHz \citep{Koen et al.1999, Billeres2005}, and no new pulsation modes were found in our data. Therefore, the range between 4.0 mHz - 8.0 mHz was plotted in Figure~\ref{fig:pulsation}. Three pulsation peaks, which are the same within the uncertainties as the previously published frequencies, were detected above the $4\sigma$ noise levels and are listed in Table~\ref{tab:pulsation}. Only the data in which these pulsations were detected above 4-$\sigma$ noise levels were used for the pulsation timing analysis.  For those pulsation peaks, day-to-day pulsation amplitude changes are observed (Figure~\ref{fig:pulsation_daily}).  Day-to-day amplitude changes were observed for other sdBV stars, such as V541 Hya, KIC 010139564, and EC 20117-4014 \citep{Randall et al.2009, Baran et al.2012,Lynas-Gray2013}. Those variations can be explained by rotational splitting, and the daily amplitude variation for AQ~Col also may be due to unresolved rotational splittings. The pulsation amplitude also changes from year to year (Fig~\ref{fig:pulsation_seasonal}). The recent ground-based data (2020 Nov - 2021 Jan) did not show any pulsations above the noise level ($\sim$ 0.3 mmag) and thus they were not usable. Transiting Exoplanet Survey Satellite ($TESS$) space telescope also observed this target in the 20-second cadence in sectors 32 (Nov 19th, 2020 - Dec 17th, 2020) and 33 (Dec 17th, 2020 - Jan 13th, 2021), a total of about 2 months. However, as shown in Fig~\ref{fig:pulsation_TESS}, the pulsations (F1, F2, and F3) were not detected above the noise level. (Note: the pulsation amplitude is expected to be quite small in the red TESS pass-band. In addition to that, according to the research of a previously known pulsating DAV star HE0532-5605, TESS cannot detect pulsations of faint, blue, compact objects if the pulsation amplitude is as small as 0.2-0.3 mmag due to the size of the telescope \citep{Bognar et al.2020}).

\figsetstart
\figsetnum{2}
\figsettitle{Power Spectra for Each Observation Run}

\figsetgrpstart
\figsetgrpnum{2.1}
\figsetgrptitle{Jan 28, 1996 to Oct 15, 2010}
\figsetplot{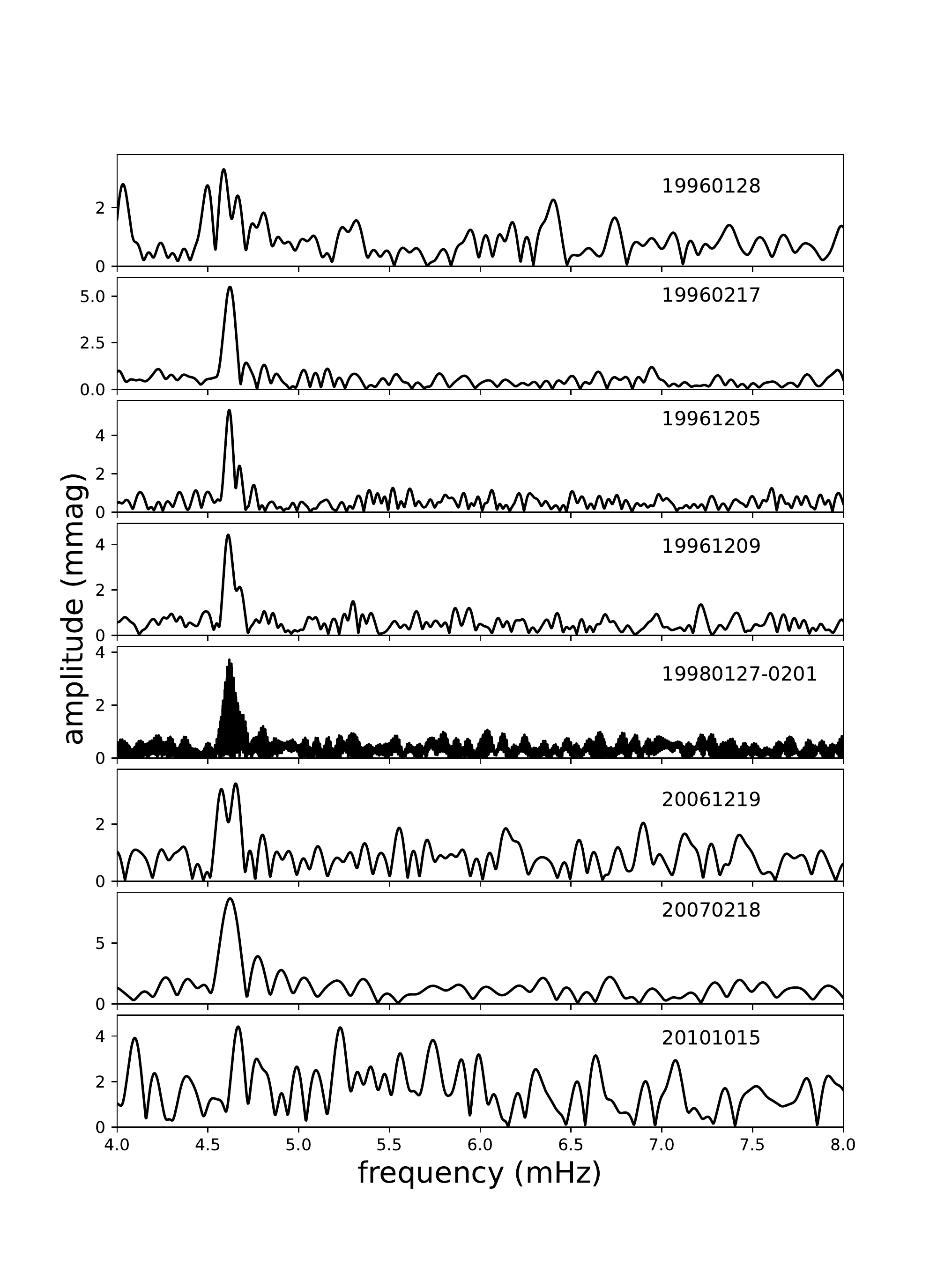}
\figsetgrpnote{Fourier Analysis}
\figsetgrpend

\figsetgrpstart
\figsetgrpnum{2.2}
\figsetgrptitle{Dec 9, 2010 to Oct 21, 2017}
\figsetplot{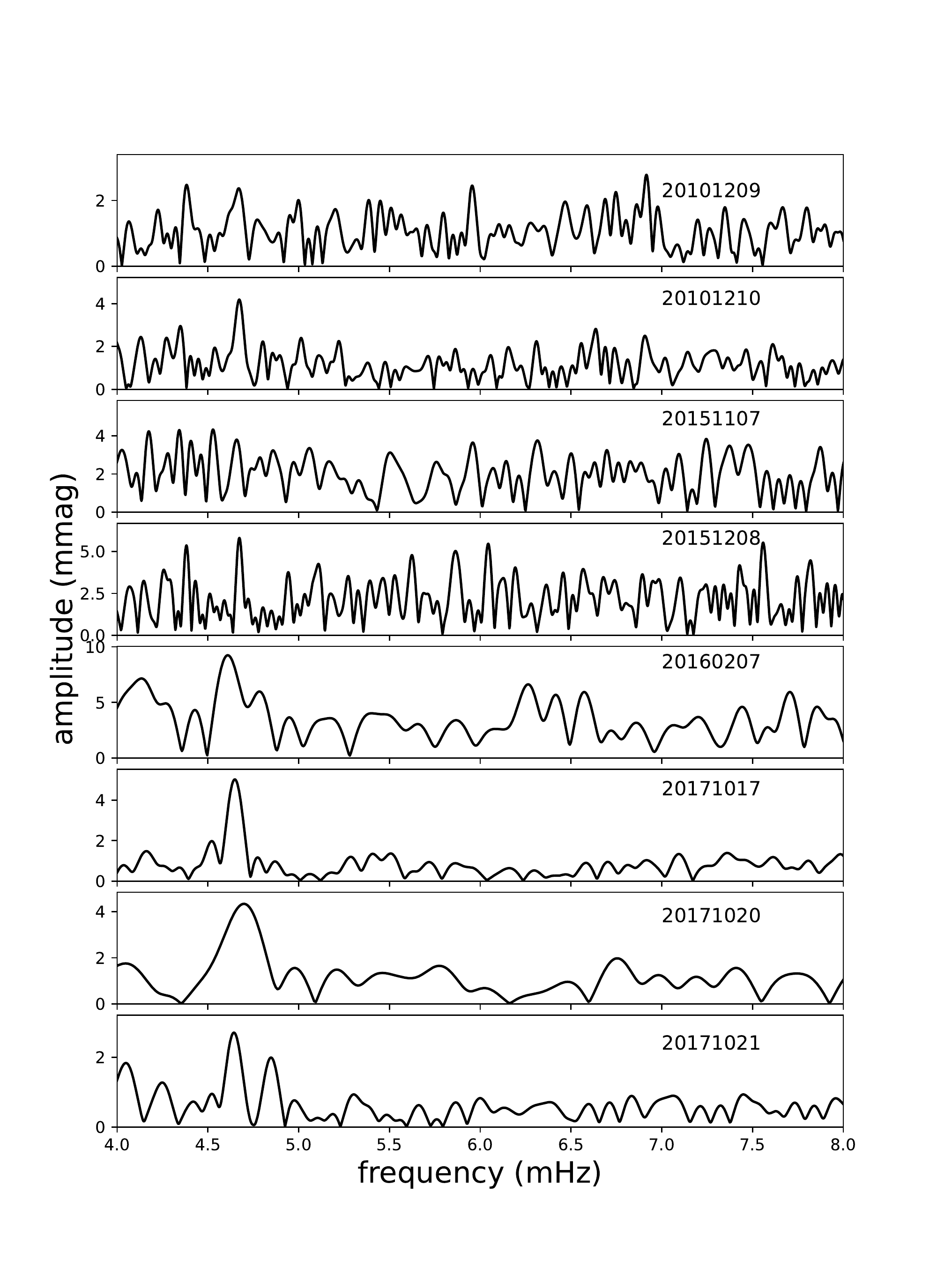}
\figsetgrpnote{Fourier Analysis}
\figsetgrpend

\figsetgrpstart
\figsetgrpnum{2.3}
\figsetgrptitle{Jan 15, 2018 to Nov 8, 2018}
\figsetplot{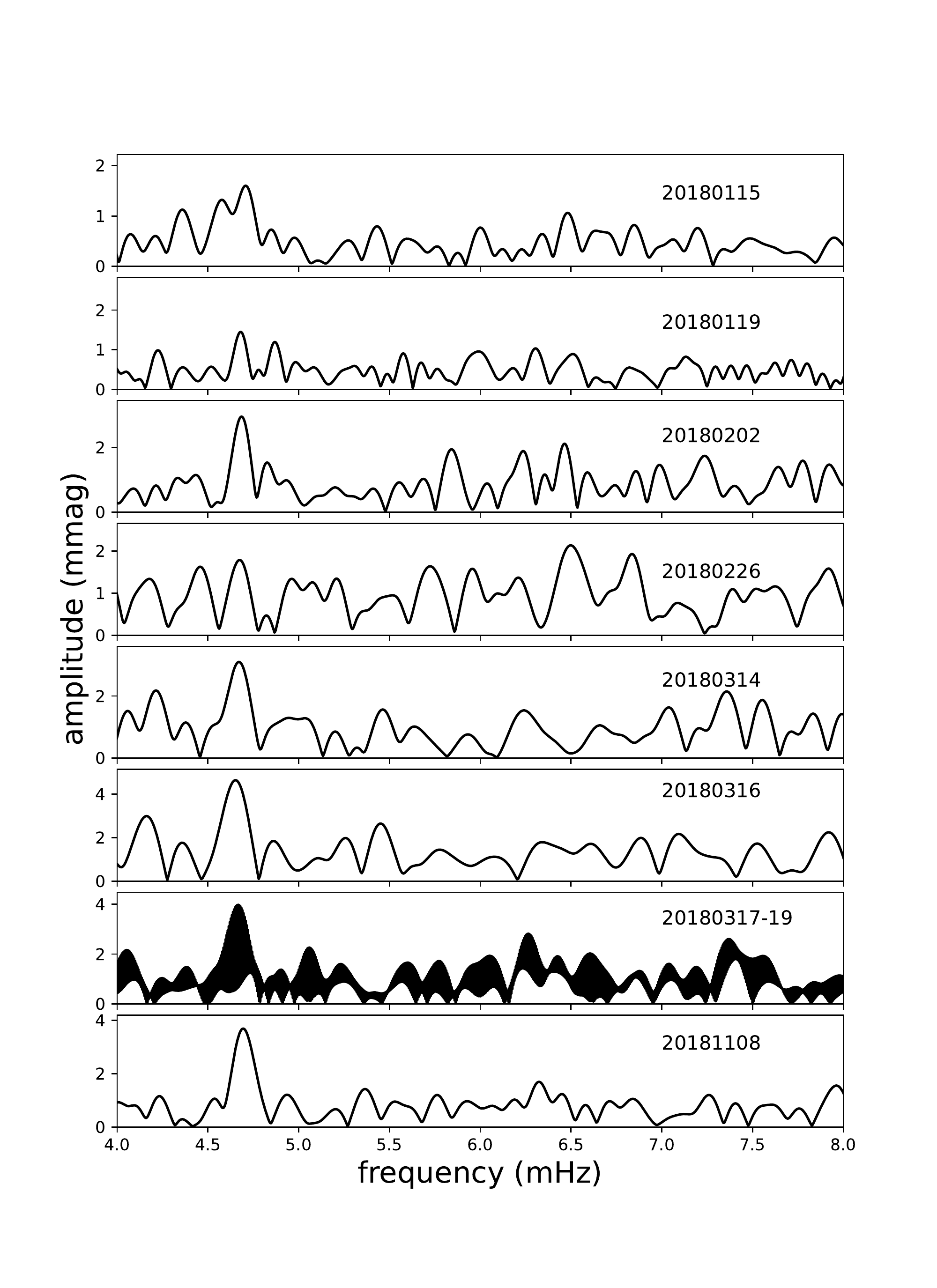}
\figsetgrpnote{Fourier Analysis}
\figsetgrpend

\figsetgrpstart
\figsetgrpnum{2.4}
\figsetgrptitle{Nov 9, 2018 to Oct 29, 2019}
\figsetplot{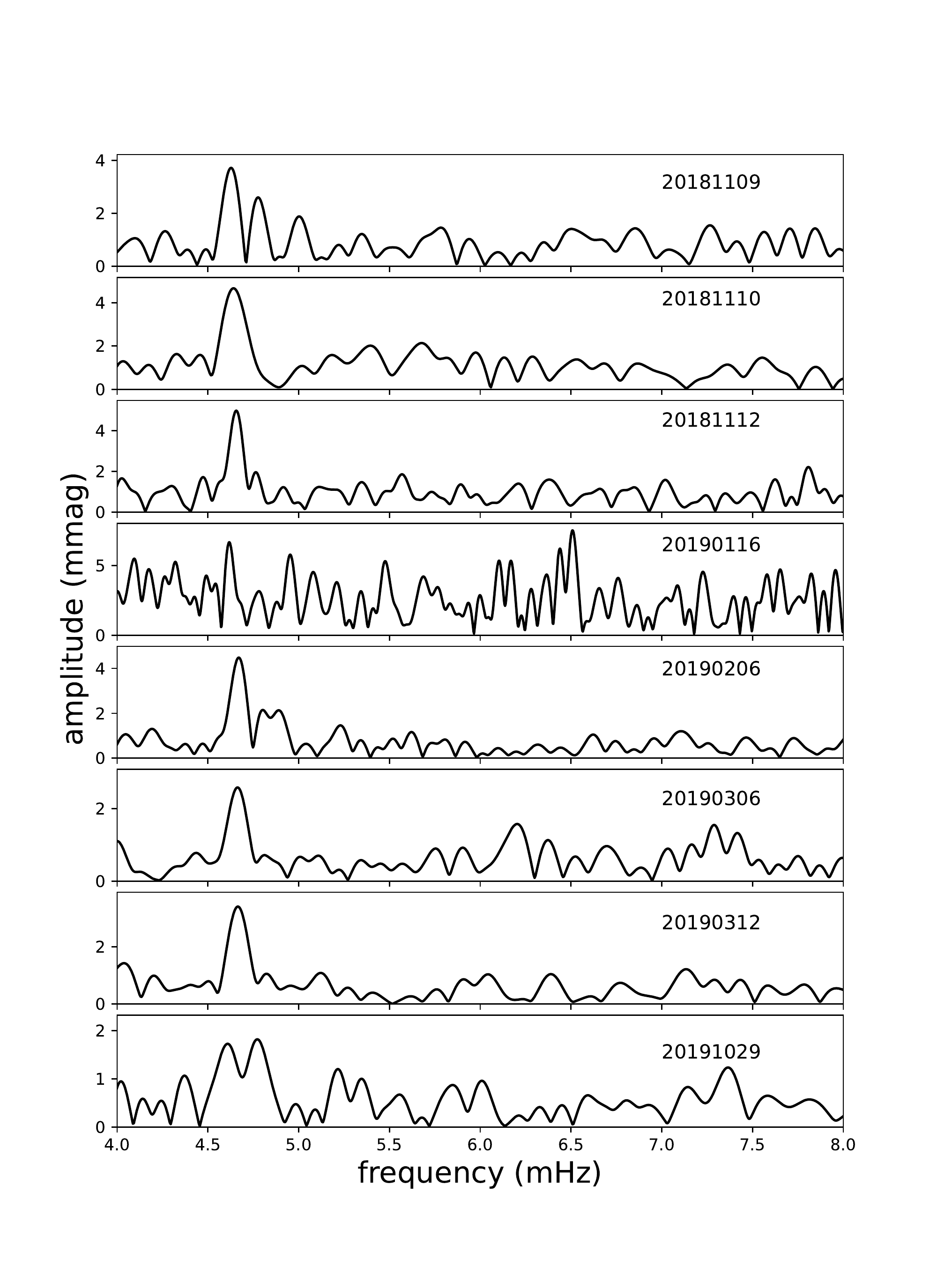}
\figsetgrpnote{Fourier Analysis}
\figsetgrpend

\figsetgrpstart
\figsetgrpnum{2.5}
\figsetgrptitle{Nov 20, 2019 to Mar 21, 2020}
\figsetplot{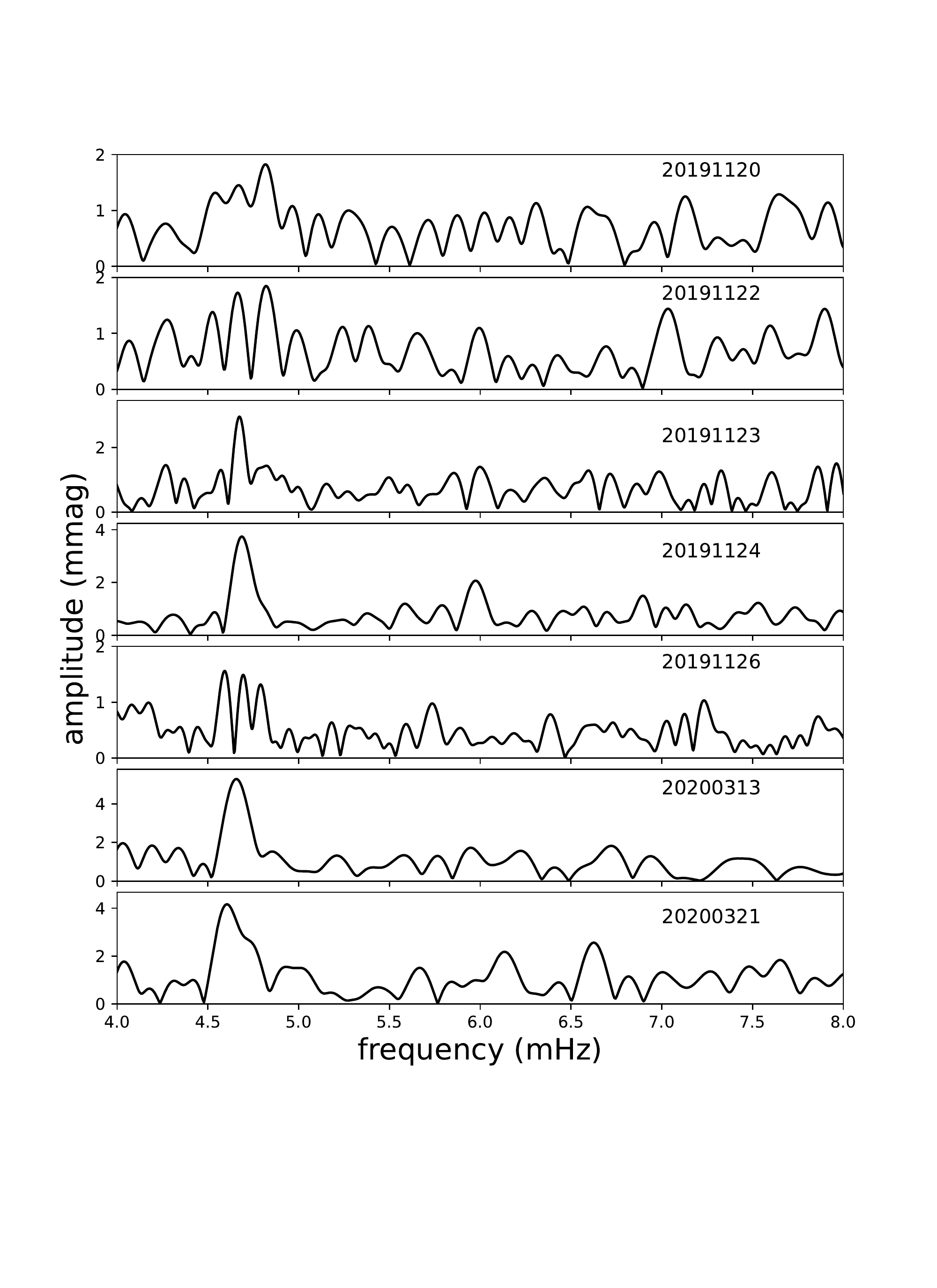}
\figsetgrpnote{Fourier Analysis}
\figsetgrpend

\figsetend

\begin{deluxetable}{cccc}
\tablecaption{Pulsation Peak Frequencies of AQ~Col  \label{tab:pulsation}}
\tablehead{\colhead{Pulsation Mode} & \colhead{Freq} & \colhead{Freq $\sigma$ } & \colhead{Period} \\ 
\colhead{} & \colhead{(mHz)} & \colhead{(mHz)} & \colhead{(s)}} 
\startdata
F1 & 4.6718377 & 1e-7 & 214.1 \\
F2 & 4.6290798 & 1e-7 & 216.0 \\
F3 & 4.5974392 & 4e-7 & 217.5 \\
\enddata
\end{deluxetable}

\subsection{Pulsation timing variation\label{subsec:OCvariation}}

Three pulsations (F1, F2, and F3) were used for the pulsation timing analysis. Data for each night were used as one data point in most cases.  However, for 1998 Jan 27 - Feb 1 and 2018 Mar 17 - 19,  two or more nights of photometry were merged to obtain better signal-to-noise (S/N) ratios. As discussed in Section~\ref{sec:OCmethod}, the secular variations -- fitted here with a quadratic -- show that the pulsation period is changing linearly due to the sdB's stellar evolution. We find $\dot{P}/P$ = $1.96 \pm 0.04 \times 10^{-7}$ yr$^{-1}$. The rate of period change ($\dot{P}$) indicates the age of the sdB star after the zero-age extreme horizontal branch (ZAEHB) \citep{Charpinet et al.2002}.  For p-modes,  $\dot{P}$ is positive during the first hot-subdwarf evolutionary phase, which is before helium in the core is exhausted.  $\dot{P}$ is negative during the second evolutionary phase, which is after the depletion of helium in its core and before the post-EHB evolution.  The change of sign occurs around 87-91 Myr after the ZAEHB according to \citet{Charpinet et al.2002} models with sdB star mass $\sim$ 0.47 $M\odot$.  However, it also depends on the hot subdwarf formation circumstances and channel. The positive values of $\dot{P}$ for \objectname[AQ Col]{AQ~Col} (sdB) thus shows that the star is still in its first evolutionary phase. The age of \objectname[AQ Col]{AQ~Col} (sdB) can also be estimated from its effective temperature, surface gravity, mass, and mass of the H-envelope.  Figure~1 of \citet{Fontaine et al.2012} also indicates that \objectname[AQ Col]{AQ~Col} is still in its first evolutionary phase.  

The relative rate of change of the radius, can also be obtained from the time scale for period change: 

\begin{equation}
\frac{\dot{P}}{P} \approx \frac{3}{2}  \frac{\dot{R}}{R}.
\label{eq:6}
\end{equation}
Here, $R$ is the radius of the star.  For  \objectname[AQ Col]{AQ~Col}, the time scale for period change ($\dot{P} / P$) calculated from F1 is $ (1.96 \pm 0.04) \times 10^{-7}$ yr$^{-1}$.  This value corresponds to the relative rate of change of the radius $\dot{R}/R \approx 1.30 \times 10^{-7}$ yr$^{-1}$.

\begin{figure}
\plotone{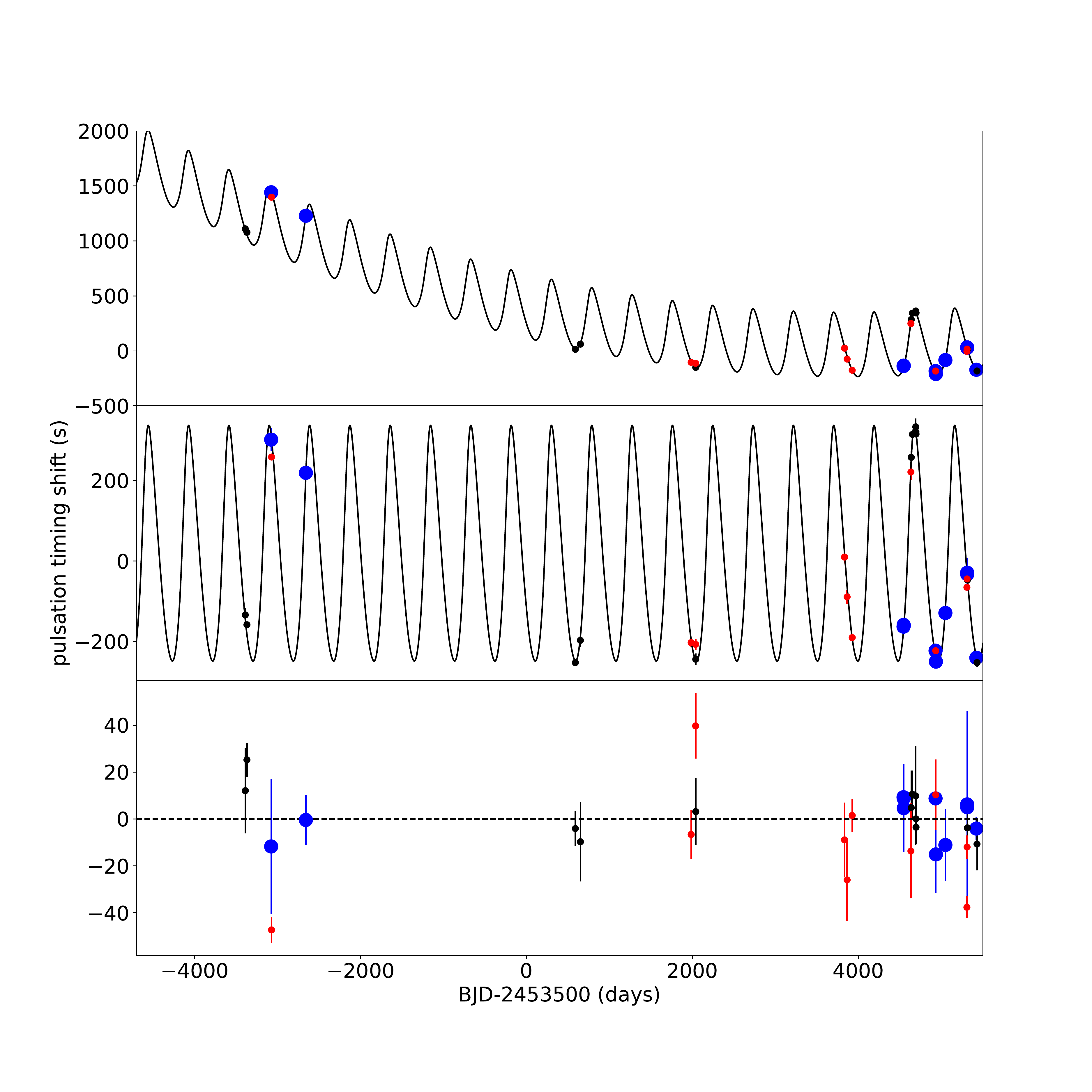} 
\caption{Top: Pulsation timing variations for \objectname[AQ Col]{AQ~Col} constructed from the F1 pulsation at 4.67 mHz. Each dot represents the pulsating timing shift of an observation run. The time scale for period change ($\dot{P} / P$) calculated from the best fit quadratic term is $ (1.96 \pm 0.04) \times 10^{-7}$ yr$^{-1}$. Middle: the top diagram after the removal of the quadratic fit using Equation~\ref{eq:binary2}. The periods and amplitudes of the fitted curve are 486.0 d and 307.8 s. Bottom: The fit curve residuals of the middle panel. The blue dots are the data points obtained from light curves with the pulsation amplitude larger than 4-$\sigma$ that were used for the fittings.  The black dots are the data points obtained from light curves with the pulsation amplitude between 3 to 4-$\sigma$. The red dots are the data points obtained from light curves with the pulsation amplitude between 2 to 3-$\sigma$.  Pulsation timing shift uncertainties of some data points in the top and middle panels are smaller than the symbol size. Most of the data points represent one day of data. However,  data of 1998 Jan 27 - Feb 1 (BJD-2453500 = -2658.7) and 2018 Mar 17 - 19 (BJD-2453500 = 4695.2) were merged to obtain better S/N ratios. \label{fig:OC}}
\end{figure}
\begin{figure}
\plotone{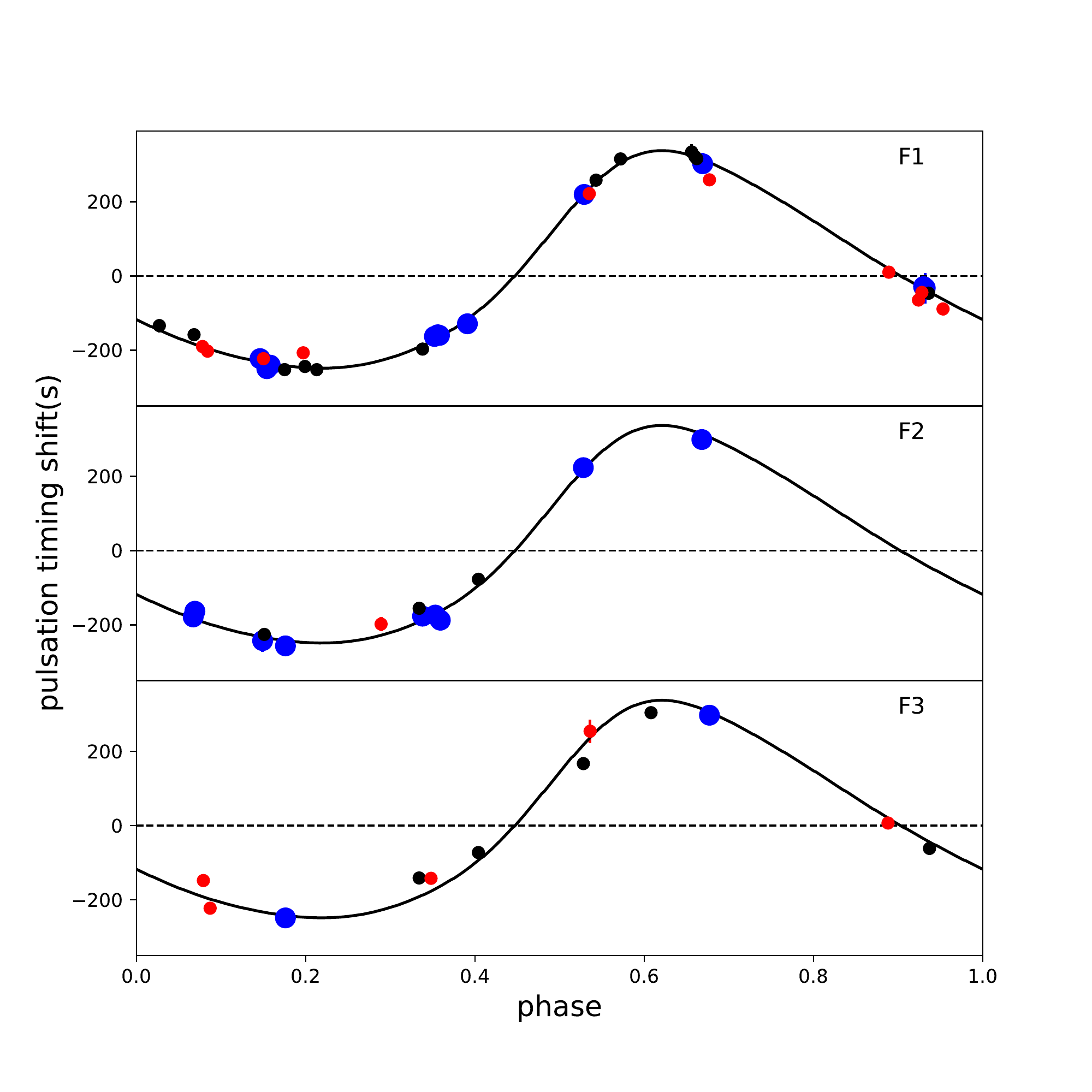} 
\caption{Phase diagram of the pulsation timing variations for \objectname[AQ Col]{AQ~Col} constructed from the F1, F2, and F3 pulsations (F1: 4.67 mHz at the top and F2: 4.63 mHz at the middle, F3: 4.60 mHz at the bottom).  Symbols have the same meaning as in Figure \ref{fig:OC}.  The orbital period implied by fitting the curve 
in Figure~7 is $486.0 \pm 0.1$~d, the corresponding semi-major axis
(in light-seconds) and orbital eccentricity being 
${a_{\rm sdB}}\,{\sin}i = 307.8 \pm 4.3$ and $e = 0.42 \pm 0.03$ 
respectively.  Those results are listed in Table~\ref{tab:orbitalinfo}}  \label{fig:OCphase} 
\end{figure}
\begin{figure}
\plotone{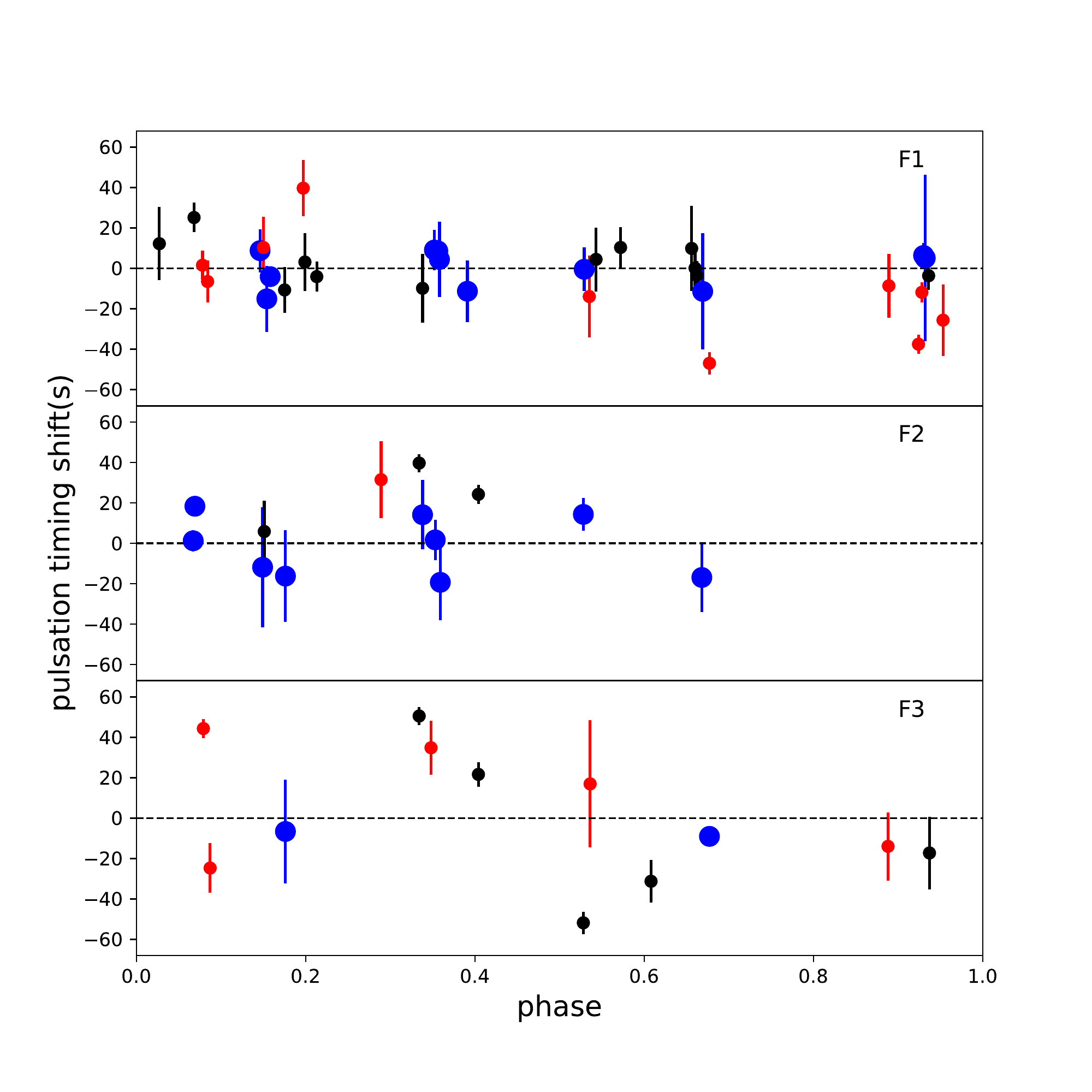}
\caption{Residuals of the pulsating timing variations after subtracting the fit shown in Figure~\ref{fig:OCphase}. The top panel shows the residuals of F1, the middle panel shows the residuals of F2, and the bottom panel shows the residual of F3.  Symbols have the same meaning as in Figure \ref{fig:OC} \label{fig:residual}}
\end{figure}

Figure~\ref{fig:OC} shows the time-series pulsating timing variation for F1 before and after the removal of the quadratic terms, and Figure~\ref{fig:OCphase} presents the phase-folded pulsation timing variation for F1, F2, and F3 after the removal of the quadratic terms.  As indicated in section~\ref{subsec:pulsation}, only the data in which the pulsation amplitude was larger than 4-$\sigma$ were used for the analysis.  However, most of the data in which the pulsation amplitude is between 2-4 $\sigma$ still match well with the solid curves in the figure, so we included those into Figures~\ref{fig:OC}, \ref{fig:OCphase}, and \ref{fig:residual} using different colors.  Table~\ref{tab:OC_phase} lists all pulsation timing shifts for the F1, F2, and F3 pulsation modes.  The solid curves in Figure~\ref{fig:OCphase} are the best fitting orbital solution (using only the F1 data in which the pulsation amplitude is larger than 4-$\sigma$) with the actual timing shifts for F1, F2 and F3 superimposed. From Table~\ref{tab:orbitalinfo}, the orbital period implied by fitting the curve 
in Figure~\ref{fig:OCphase} is $486.0 \pm 0.1$~d, the corresponding semi-major axis
(in light-seconds) and orbital eccentricity being 
${a_{\rm sdB}}\,{\sin}i = 307.8 \pm 4.3$ and $e = 0.42 \pm 0.03$ 
respectively. The orbital solutions are shown in Table~\ref{tab:orbitalinfo}. The formal $\chi ^2$ values (only using the data in which the pulsation amplitudes are larger than 4-$\sigma$) are 12.8 (F1) and 5.3 (F2).  The degrees of freedom of F1 and F2 are 13 and 6.  We did not calculate $\chi ^2$ for F3 because only two data points have pulsation amplitudes above 4-$\sigma$.  The corresponding right tail p-values are 0.54 and 0.49.  Therefore, model fits are acceptable. The $\chi ^2$ values being consistent with the number of degrees of freedom suggest that all relevant physical information has been extracted from the data. 

The mass function (Equation (\ref{eq:massfunction})), as computed from the pulsation timing variation, is $f$ = 0.133 $\pm$ 0.006 $M_{\odot}$. Assuming this system consists of only the sdB and long orbital period binary (please see the following sections for the triple stars possibility), for a canonical sdB star mass of $M_1 = 0.5{\,}{\rm M}_{\odot}$, 
substituting $f$ and $M_1$ into Equation~(\ref{eq:massfunction}) gave
$M_2 = 0.515 \pm 0.012\,{\rm M}_{\odot}$ and
$M_2 = 0.645 \pm 0.016\,{\rm M}_{\odot}$ for 
$i$ = 90$^{o}$ and $i$ = 60$^{o}$ respectively.
The estimated amplitude of the RV variations is $K_{sdB}$ = 15.2 $\pm$ 0.3 km/s. Although radial velocities are difficult to measure for sdB stars because of their high gravity (which broadens the line profiles), \citet{Silvotti et al.2020} succeeded in measuring the radial velocities of sdB stars to a precision of $\approx$ 100 m/s (5-sigma level) using Harps-N at the 3.6~m $Telescopio\ Nazionale\ Galileo$ (TNG).  The radial velocity amplitude of AQ Col due to the existence of the long orbital period binary is much larger than this, and it should be possible to confirm the companion using the radial velocity method. 


\clearpage

\startlongtable
\begin{deluxetable}{ccccccc}
\tablecaption{Pulsation timing variation data for F1 (4.67 mHz), F2 (4.63 mHz), and F3 (4.60 mHz) pulsation modes after removing the quadratic fits. All data points obtained from light curves with the pulsation amplitudes larger than 2-$\sigma$ are listed. \label{tab:OC_phase}}
\tablehead{\colhead{Time} & \colhead{F1} & \colhead{$\sigma$ (F1)} & \colhead{F2} & \colhead{$\sigma$} (F2)& \colhead{F3} & \colhead{$\sigma$ (F3)} \\ 
\colhead{(BJD-2453500)} & \colhead{(s)} & \colhead{(s)} & \colhead{(s)} & \colhead{(s)} & \colhead{(s)} & \colhead{(s)} } 

\startdata
-3388.61 & -133.9                              & 18.2                   &                                     &                        &                                     &                        \\
-3368.71 & -158.3                              & 7.3                    & -178.4                              & 5.4                    &                                     &                        \\
-3367.63 &                                     &                        & -163.0                              & 4.2                    &                                     &                        \\
-3076.56 & 301.9                               & 28.7                   & 299.3                               & 17.2                   &                                     &                        \\
-3072.57 & 258.6                               & 5.6                    &                                     &                        & 297.0                               & 4.1                    \\
-2658.69 & 219.3                               & 10.7                   & 223.7                               & 8.2                    & 166.7                               & 5.6                    \\
589.39   & -252.5                              & 7.5                    &                                     &                        &                                     &                        \\
650.33   & -197.1                              & 16.9                   & -176.2                              & 17.1                   &                                     &                        \\
1984.79  & -202.7                              & 10.4                   &                                     &                        &                                     &                        \\
1985.69  &                                     &                        &                                     &                        & -222.6                              & 12.3                   \\
2039.70  & -207.2                              & 13.9                   &                                     &                        &                                     &                        \\
2040.69  & -244.0                              & 14.3                   &                                     &                        &                                     &                        \\
2112.61  &                                     &                        &                                     &                        & -142.1                              & 13.3                   \\
3832.65  &                                     &                        &                                     &                        & 6.7                                 & 16.9                   \\
3833.74  & 9.9                                 & 15.9                   &                                     &                        &                                     &                        \\
3864.70  & -89.0                               & 17.7                   &                                     &                        &                                     &                        \\
3925.70  & -190.2                              & 7.2                    &                                     &                        & -148.1                              & 4.8                    \\
4182.77  &                                     &                        &                                     &                        & 303.8                               & 10.4                   \\
4544.56  & -163.1                              & 10.0                   & -173.3                              & 10.1                   &                                     &                        \\
4546.52  & -158.6                              & 1.7                    &                                     &                        &                                     &                        \\
4547.55  & -160.2                              & 18.7                   & -187.1                              & 18.9                   &                                     &                        \\
4633.72  & 221.4                               & 20.2                   &                                     &                        & 254.0                               & 31.5                   \\
4637.64  & 257.8                               & 15.8                   &                                     &                        &                                     &                        \\
4651.61  & 314.9                               & 10.0                   &                                     &                        &                                     &                        \\
4692.29  & 333.7                               & 21.1                   &                                     &                        &                                     &                        \\
4694.28  & 320.9                               & 10.4                   &                                     &                        &                                     &                        \\
4695.24  & 315.7                               & 7.3                    &                                     &                        &                                     &                        \\
4930.55  & -222.8                              & 10.7                   &                                     &                        &                                     &                        \\
4931.51  &                                     &                        & -242.2                              & 29.7                   &                                     &                        \\
4932.55  & -222.9                              & 15.1                   & -225.4                              & 15.3                   &                                     &                        \\
4934.53  & -249.9                              & 16.3                   &                                     &                        &                                     &                        \\
4999.70  &                                     &                        & -197.6                              & 19.1                   &                                     &                        \\
5021.34  &                                     &                        & -155.1                              & 4.4                    & -141.2                              & 4.4                    \\
5049.31  & -129.0                              & 15.3                   &                                     &                        &                                     &                        \\
5055.30  &                                     &                        & -77.0                               & 4.8                    & -72.8                               & 6.1                    \\
5308.51  & -65.1                               & 4.7                    &                                     &                        &                                     &                        \\
5310.46  & -44.5                               & 5.0                    &                                     &                        &                                     &                        \\
5311.50  & -29.0                               & 6.3                    &                                     &                        &                                     &                        \\
5312.48  & -32.8                               & 41.1                   &                                     &                        &                                     &                        \\
5314.51  & -46.7                               & 7.1                    &                                     &                        & -61.6                               & 18.0                   \\
5422.30  & -240.4                              & 4.8                    &                                     &                        &                                     &                        \\
5430.28  & -252.4                              & 11.3                   & -256.3     \\
\enddata
    \tablecomments{Time is mid-observing time.}
    \tablecomments{Table 4 is published in its entirety in the machine-readable format.}
\end{deluxetable}

\begin{deluxetable}{lc}[h]
\tablecaption{Orbital information of the AQ~Col system \label{tab:orbitalinfo}}
\tablehead{\colhead{Parameters} & \colhead{values} \\ 
\colhead{} & \colhead{} } 
\startdata
Period, P (days) & 486.0 $\pm$ 0.1 \\
Amplitude, $a_{sdB} \sin{i}$ (s) & 307.8 $\pm$ 4.3 \\
Eccentricity, $e$ & 0.42 $\pm$ 0.03 \\
Argument of periapsis, $\varpi$ (rad) & 0.72 $\pm$ 0.05 \\
Zero point of time, $t_0$ (BJD-2453500) & 262.5 $\pm$ 3.6 \\
Mass function, $f$ ($M_{\odot}$) & 0.133 $\pm$ 0.006 \\
Radial velocity for sdB star, $K_{sdB}$ (km/s) & 15.2 $\pm$ 0.3 \\
\enddata
\end{deluxetable}

When the period of the pulsation timing variation matches with the pulsation $amplitude$ variation, the pulsation timing variation may be due to two closely spaced pulsation frequencies \citep{Lutz2011}. However, this is not the case. The period of the pulsation timing variation (P= 486.0 d) does not match with the F1, F2 pulsation amplitude variations and the pulsation amplitude variations' shapes are not sinusoidal, so this is not due to the beating of two closely spaced pulsation frequencies (see Figure~\ref{fig:pulsation_seasonal}).  Therefore, we conclude that the resulting pulsation timing variations in Fig~\ref{fig:OC}~and~\ref{fig:OCphase} are due to the light-travel effects caused by an unseen companion. 

 \subsection{Spectroscopy\label{subsec:spectroscopy}}

As noted in Table~\ref{tab:number5}, AQ~Col was found to have a large
and rapid radial velocity variation.  In particular, spectra obtained
on 1996 December $5^{\rm th}$ showed a radial velocity change of
$\sim 50\,$km/s in 45~minutes.  An orbital period of 486~days,
identified through photometry, corresponds to a sdB star radial velocity
amplitude of $15.2 \pm 0.3\,$km/s (Table~\ref{tab:orbitalinfo}), smaller than
the radial velocity change observed on 1996 December $5^{\rm th}$.
We have therefore considered AQ~Col to be a triple star, the long-period
binary identified photometrically having a sdB component which is a
short-period binary, and proceeded to analyse our reduced spectra
on this basis.

During four 1200-s exposures, resulting in the two spectra obtained on
1996 December $5^{\rm th}$, orbital motion would in this case have
resulted in significant Balmer line broadening.  We therefore decided to
confine our spectroscopic analysis to the two spectra obtained in
1996 December; these were shifted into an observer's rest-frame and
added with equal weight.  Balmer line broadening by orbital motion was
found to be well-represented by convolution with a Gaussian having a 
${\rm FWHM} = 0.5\,$~\AA.  In determining the equivalent FWHM, 
a time-dependent linear change in radial velocity (sampled at 0.1-s
intervals) was assumed during the sequence of arc and science exposures.

\begin{figure}
\includegraphics[width=150mm]{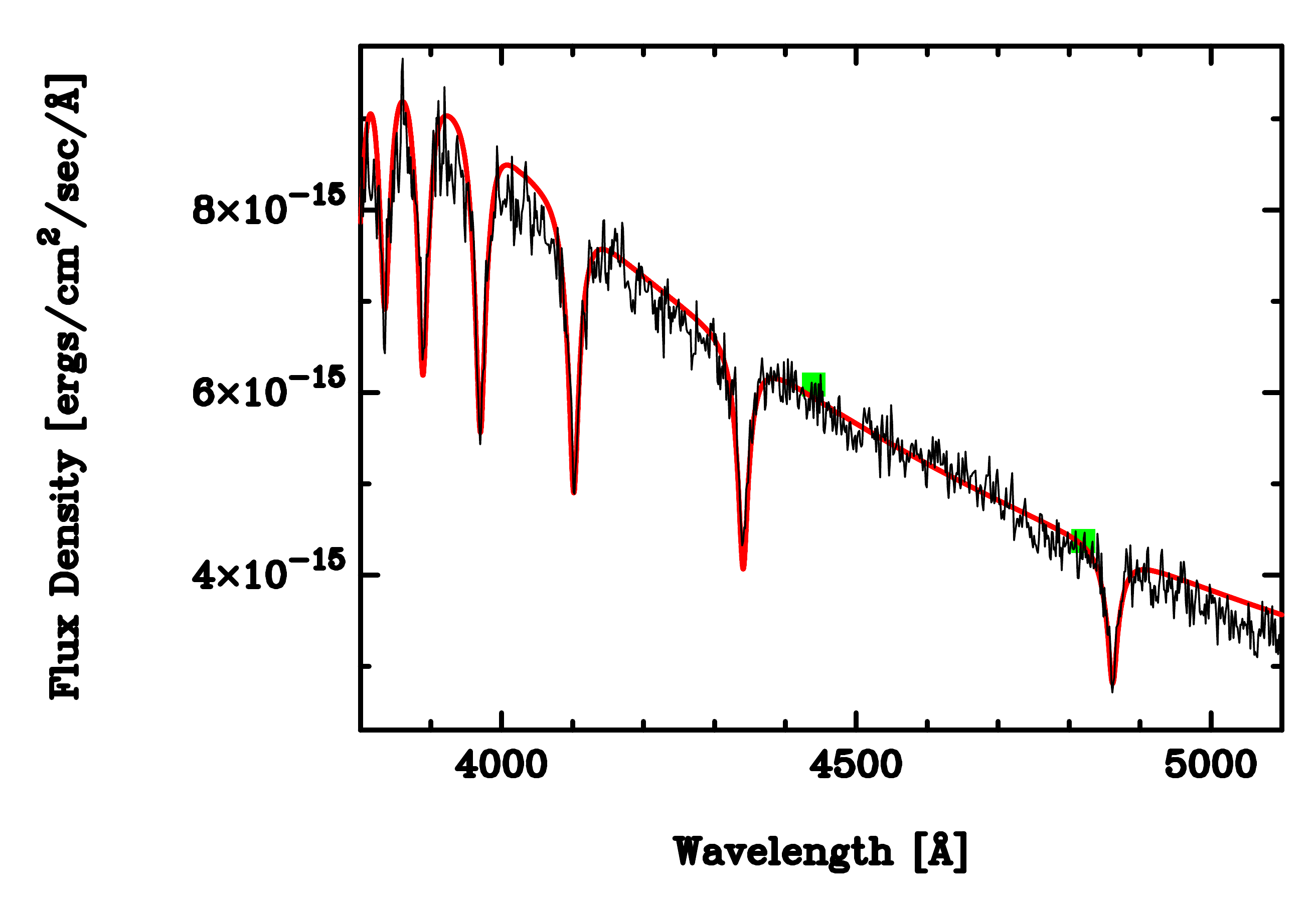}
\caption{Two spectra obtained on 1996 December $5$ combined after shifting into an observer's rest frame, compared with a non-LTE model spectrum for $T_{\rm eff} = 30000$K, $\log g = 5.9$, $\log (N({\rm He})/N({\rm H})) = -5.0$ and $v{\sin}i = 300\,$. Green squares show the dereddened Johnson B and Gaia G band flux
densities for comparison.
See text for details. \label{fig:fit}}
\end{figure}

Near-photometric conditions prevailed at the SAAO Sutherland site on
1996 December $5^{\rm th}$, and a small wavelength-dependent correction
was applied using available photometry.  UBV fluxes were obtained from
the EC~Survey \citep{2013MNRAS.431..240O} and zero-points of 20.94,
20.51 and 21.12 for U, B and V respectively; these being derived from
the \citet{1975ApJ...197..593H} flux calibration and the 
\citet{1979ApJS...40....1K} model atmosphere for Vega.  While the
UBV flux calibration is dated, it proved to be consistent with the
Gaia G-Band flux density \citep{2021A+A...649A...3R} and adequate for
our purposes, given other uncertainties involved.

\citet{2021AJ....161..147B} obtain $1.54$ $^{+0.10}_{-0.07}$ ~kpc as the
AQ~Col distance, with its galactic coordinates 
($\ell$ = 243$^o$.84,b = -33$^o$.84),
the \citet{1992A+A...258..104A} galactic extinction map gives
$E_{{\rm B}-{\rm V}} = 0.06 \pm 0.05$, assuming $R = 3.1$. Adopting $E_{{\rm B}-{\rm V}} = 0.06$, both the spectral
energy distribution and flux density points have been
dereddened using \citeauthor{1979MNRAS.187P..73S}'s
\citeyearpar{1979MNRAS.187P..73S} calibration and 
\citeauthor{1983MNRAS.203..301H}'s \citeyearpar{1983MNRAS.203..301H} 
extension of it into the optical and infrared regions.
Figure~\ref{fig:fit} shows the dereddened AQ~Col energy distribution 
obtained from 1996 December spectra plotted as a black line. 
Green squares show the dereddened Johnson B and Gaia G band flux
densities for comparison. 

Spectra obtained on 1989 December $21^{\rm st}$ and 
1996 December $5^{\rm th}$ were ``quick look'' spectra obtained with 
a Reticon detector and taken for classification purposes as part of the
Edinburgh-Cape Survey \citep{Stobie et al.1997}.  No special 
attention was accordingly given to the removal of ``fixed pattern'' 
noise known to be present in Reticon spectra as, for example, 
\citet{1983PASP...95..810T} discusses. A comparison between the 
combined spectrum plotted in Figure~9 and the AQ~Col spectrum 
\citet[their figure 6]{Koen et al.1999} publish, 
demonstrates a probable non-astrophysical origin for the apparent 
structure in the Figure~9 spectrum.  We accordingly analyze the 
Figure~9 spectrum on the basis that only Balmer lines are present, 
as indicated by charge-coupled-device (CCD) spectra, one obtained by 
ourselves with the SOAR~4.1-m telescope and the other which 
\citeauthor{Koen et al.1999} publish.

The red line in Figure~\ref{fig:fit} is the non-LTE model atmosphere
emergent energy distribution from the \citet{2014ASPC..481...95N} 
grid for $T_{\rm eff} = 30000$K, $\log g = 5.9$ and 
$\log (N({\rm He})/N({\rm H})) = -5.0$, successively broadened by two
Gaussians having ${\rm FWHM} = 0.5\,$~\AA\, and ${\rm FWHM} = 3.5\,$~\AA\,
to allow for orbital and instrumental broadening respectively.
Trial rotation-broadening profiles were calculated for selected
$v\,{\sin}\,i$ values following  (for example)
\citet{1968BAN....19..526U} and adopting a grey atmosphere 
limb-darkening; the result was convolved with the synthetic spectrum
for the fitted effective temperature and surface gravity for 
subsequent comparison with observation.  At our low resolution, 
Balmer line cores are not very sensitive to stellar rotation and we 
therefore have a large error in the determination of 
$v\,{\sin}\,i = 300 \pm 100 \,$(km/s). The Figure~\ref{fig:fit}
model spectrum has therefore also been broadened by a rotation profile
corresponding to $v{\sin}i = 300\,$km/s.  Our model spectrum plotted
in Figure~\ref{fig:fit} is the best agreement with observation we
achieved; other comparisons gave standard error limits of 
$\delta(T_{\rm eff}) = \pm 2000$K and $\delta(\log g) = \pm 0.2$.
The absence of He~I lines and the good fit obtained 
with $\log (N({\rm He})/N({\rm H})) = -5.0$ indicates that this may be
a helium abundance upper limit.

Synthetic spectrum fitting involved normalisation to the Johnson
V-Band flux at 5540 \AA, corresponding to a hot subdwarf angular radius
of $\alpha = (2.6 \pm 0.1) \times 10^{-12}$~radians. The uncertainty in
$\alpha$ follows from the $T_{\rm eff}$ and reddening correction errors.
Given the \citeauthor{2021AJ....161..147B} distance, the hot subdwarf
radius, mass and luminosity were then found to be 
$0.18 \pm 0.01\,{\rm R}_{\odot}$,
$0.91 \pm 0.44\,{\rm M}_{\odot}$ and
$24 \pm 7\,{\rm L}_{\odot}$ respectively.
Associating the measured angular radius with a stellar radius implies
that AQ~Col is a spherical star; with $v{\sin}i = 300\,$km/s this may
not be the case and could have led to an erroneous high mass.

 \subsection{Unseen companions\label{subsec:companion}}
   Most subdwarf-B (sdB) stars in binary systems have companions which
are white dwarfs or M-dwarf main sequence stars
\citep{2015A&A...576A..44K}; these have short 
orbital periods $(\lesssim 10\ {\rm days})$ and are believed to be post-common envelope systems \citep{Han et al.2002,Han et al.2003,2017A&A...599A..54X}.
Some sdB binaries have longer orbital periods with an F- or G-type giant or
main sequence star and 26 of those systems have been studied so far \citep{Vos2017, Vos2019a}. 
Following \citet{Han et al.2002,Han et al.2003}, sdB stars in long period binaries are formed as a consequence of a red giant progenitor losing almost all of its hydrogen-rich envelope, at the onset of core helium-burning, through stable Roche lobe overflow (RLOF);
their calculations suggest that the orbits should be circular and
have periods $\lesssim 500\ {\rm days}$.
However, radial velocity observations by \citet{2012ASPC..452..163O, Deca et al.2012, Barlow et al.2013, Wade et al.2014}, identified sdB stars having a main-sequence or giant companion with orbital periods $> 500\ {\rm days}$, significantly greater
than the \citet{Han et al.2002,Han et al.2003} orbital-period
distribution would suggest.
\citet{2013MNRAS.434..186C} reproduce the orbital-period distribution observed by
\citet{2012ASPC..452..163O} using detailed binary evolution 
calculations for the stable RLOF channel, improving on the simplified
binary population synthesis by \citet{Han et al.2003}. \citet{Vos2017, Vos2020} also performed a binary population synthesis study, and  the estimated period of binaries that went through this RLOF channel is P = 400 - 1500 d, and eccentricity e = 0 - 0.3 (See Figure 2 of \citet{Vos2019b} and Figure 3 of \citet{Molina et al.2021}). The orbital period of the long orbital period companion to \objectname[AQ Col]{AQ~Col} (P = 486.0 d) falls in the middle of this range, however the eccentricity (e = 0.424) does not fall in the range. That suggests that this system in not a typical sdB+MS binary system.  This high eccentricity might be caused by the presence of a close companion which was discussed in section \ref{subsec:spectroscopy}.  The common envelope system might cause a large amount of mass loss from the system that the outer companion could not efficiently accrete and its orbit was not efficiently circularized.  If AQ~Col is eventually confirmed to be a triple-star as proposed in the
present paper, the two-star RLOF model needs an adaptation dependent on
better component mass determinations.  Triple star
evolution is an active field of research lagging behind studies of
binary and single star evolution as \citet{2020A+A...640A..16T} discuss.
An interesting result from the \citeauthor{2020A+A...640A..16T} study
is the fact that about 40\% of triple-star systems retain eccentric 
orbits, before RLOF if it occurs, which would be consistent with our 
determination of $e = 0.42 \pm 0.03$ for the orbit of the unseen more 
distant component.

\begin{figure}
\plotone{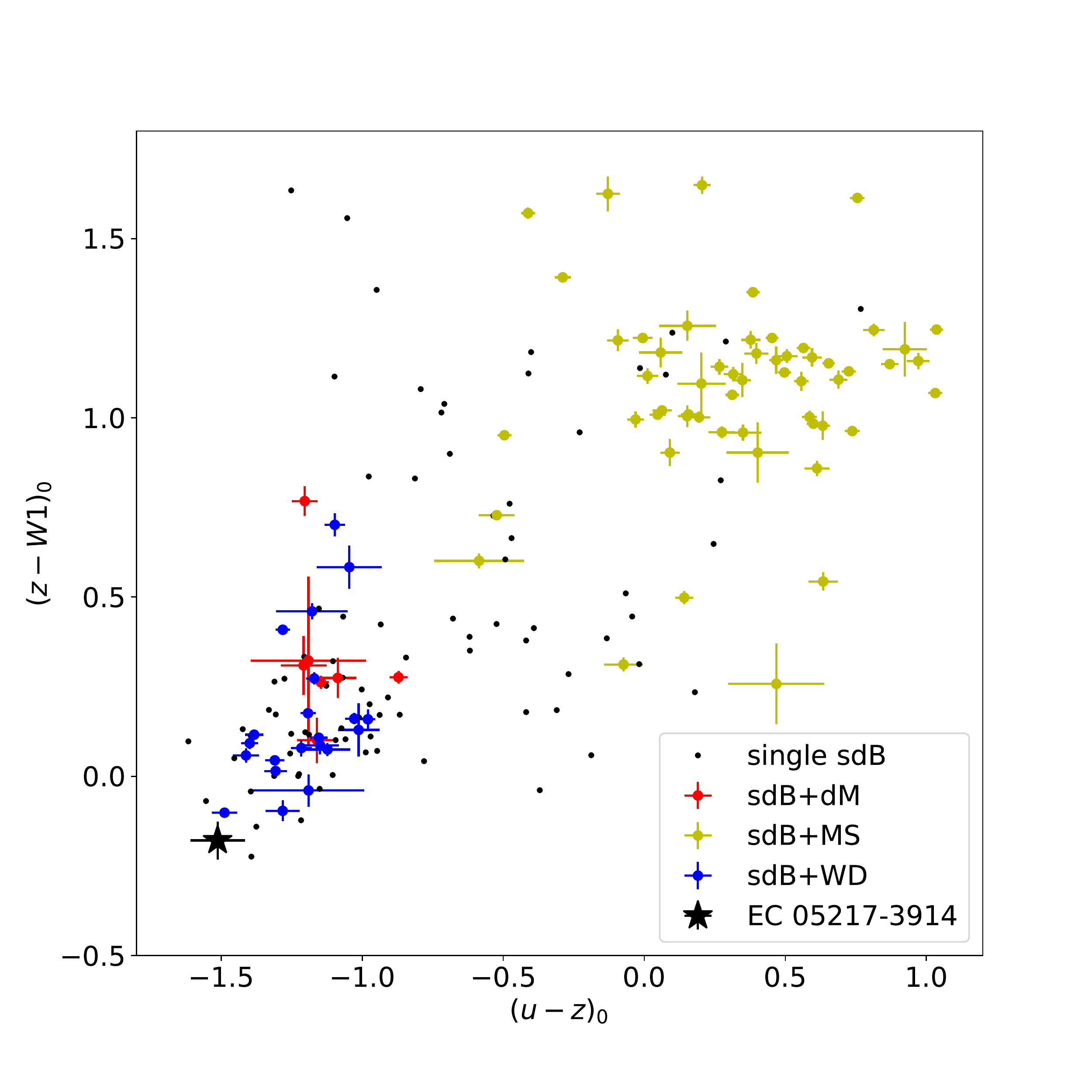}\\
\caption{Skymapper/Wise u-z vs. z-W1 color-color diagram (colors dereddened). The black, red, yellow, and blue symbols show the single sdB stars, sdB+dM, sdB+MS, sdB+WD, respectively. AQ~Col sdB binary candidate is indicated with a black star symbol. All stars except single sdB stars have uncertainties which are indicated by the cross marks. Those empirical data are taken from all available data in the subdwarf database\citep{Geier et al.2019}.  \label{fig:color}}
\end{figure}

Figure~\ref{fig:color} shows the color-color diagram (u-z vs. z-W1) of known sdB binary systems. The Skymapper u and z and WISE W1 magnitudes are used \citep{Keller et al.2007, Wright et al.2010}. In this diagram, the colors of AQ~Col are compared with single sdB stars, sdB+MS, sdB+dM, and sdB+WD.  All the available data used are from the hot subdwarf database \citep{Geier et al.2019}.  This diagram suggests that the AQ~Col system is not likely to be a sdB+dM or sdB+MS system and the color is bluer than other known sdB+WD systems. The known single sdBs spread even into the sdB+MS regions, and those might actually be binary systems. 

This result is the same for the GAIA color. The GAIA color $G_{BP} - G_{RP} = -0.435$ and absolute magnitude ${\rm M}_{\rm G} = 4.52$ for AQ~Col,
neglecting a small reddening correction, places it among the 
apparently single sdB and sdOB stars in the color-magnitude diagram 
\citet[their figure 2, right-hand panel]{Geier et al.2019} publish
and verifying their suggestion that some of these could also be binary 
(or multiple) systems.

As discussed in the Section~\ref{subsec:spectroscopy}, Spectra obtained on 1996 December $5^{\rm th}$ exhibit a change of 
49.1 km/s in 46.1 minutes which cannot be a consequence of the 
wide-binary inferred through the light-travel-time analysis.  Instead we
suggest the hot subdwarf in the AQ~Col wide-binary is itself
a close-binary.  If the close-binary orbit were circular, coplanar
with the line-of-sight and the hot subdwarf had an orbital speed about 
this center-of-mass of $\sim 220$ km/s, as Table~\ref{tab:number5} 
spectra from 1989 and 2020 seem to suggest,
an unseen companion would have a mass 
of $\sim 1.4{\,}{\rm M}_{\odot}$ for a canonical hot subdwarf mass of 
$\sim 0.5{\,}{\rm M}_{\odot}$; a systemic velocity of zero has been
adopted, the barycentric correction relative to the Solar System 
barycentre is $\sim 0.1$ km/s and has been neglected.  If the
hot subdwarf orbital speed were 300 km/s, which could be the case if
an orbit were highly inclined, the companion mass would be 
$\sim 3.0{\,}{\rm M}_{\odot}$. Estimated orbital periods range from 0.7
to 1.1 days for the 220 and 300 km/s cases respectively.  Varying the
assumed systemic velocity by $\pm 10$ km/s and the canonical hot 
subdwarf mass by $\pm 0.1{\,}{\rm M}_{\odot}$ alters the estimated 
companion mass by $\sim 0.05{\,}{\rm M}_{\odot}$ and
$\sim 0.1{\,}{\rm M}_{\odot}$ respectively. If the AQ~Col sdB star in a wide-binary itself forms a close binary
with a $1.4\,{\rm M}_{\odot}$ companion (for example) then with
$f = 0.133 \pm 0.006\,{\rm M}_{\odot}$ 
and $M_1 = 1.9{\,}{\rm M}_{\odot}$, 
substituting $f$ and $M_1$ into Equation~(15) gave wide-binary
companion masses of
$M_2 \simeq 1.05  \pm 0.02\,{\rm M}_{\odot}$ and
$M_2 \simeq 1.27  \pm 0.02\,{\rm M}_{\odot}$ for 
$i = 90^{\rm o}$ and $i = 60^{\rm o}$ respectively. If the inclination is below $i = 53.8^{\rm o}$, the mass of the wide-binary would be larger than the Chandrasekhar limit ($M_2 =  1.40 {\rm M}_{\odot}$).

To date, several circumbinary planetary systems or brown dwarf systems candidates are discussed using the eclipse timing method (\citet{Pulley2018}, and references therein). However those long period companions have smaller O-C amplitudes than the long orbital period companion to AQ~Col, which indicates that the long orbital period companion to AQ~Col should have much larger mass. For this reason, AQ~Col may be an interesting triple star system candidate to continue monitoring - if pulsation persists.  \citet{Wu2018, Wu2020} presented mass-transfer processes from a primordial binary that evolves into an sdB+neutron star system. This theory estimates about 7000 - 21000 sdB+NS binaries in the Galaxy at the present epoch, which contributes 0.3-0.5 \% of the total sdB binaries, but no sdB+NS binary system is known yet. 


  \section{Conclusions}\label{sec:conc}

This paper discusses the photometric and spectroscopic properties of the sdB star, AQ~Col.   Photometric data of AQ~Col for 25 years show obvious periodic variations in the three largest amplitude pulsation frequencies and allowed us to obtain an orbital solution for the long period companion using the pulsation timing method. We find this long orbital period system to have the following properties:
\begin{enumerate}
\item Orbital period P = 486.0 $\pm$ 0.1  days
\item Eccentricity e = 0.42 $\pm$ 0.03, which is extremely high compared with other sdB+MS binary systems, and suggests that this system is not a typical sdB+MS binary. 
\item The light-travel time amplitude A = 307.8 $\pm$ 4.3 s
\item The expected radial velocity amplitude of the sdB star due to this companion is $K_{sdB}$ = 15.2 $\pm$ 0.3 km/s
\end{enumerate}

However, available spectra show the minimum radial velocity amplitude is $\sim$ 300 km/s, which cannot be reconciled with the estimated radial
velocity amplitude ($K_{sdB}$ = 15.2 $\pm$ 0.3 km/s). This discrepancy suggests that AQ~Col may be a triple system with both a long period (P = 486 days) and a possible short period (P $\leq$ 10 d) companion, the latter of which is below the detection limits using the pulsation timing analysis.  Our color-color diagram shows that one of the companions is likely to be a white dwarf or another hot and faint object. Since those systems have not been studied well yet, AQ~Col is a unique system that should be monitored in the future.

\citet{Pelisoli et al.2021} find HD 265435 to be a possible 
supernova la progenitor, and AQ Col could be similar.  Further AQ Col radial
velocity observations are needed to confirm that this wide 
binary has a hot subdwarf which itself is a close binary
whose components have a combined mass that exceeds
the Chandrasekhar-Mass.
\newpage
\acknowledgements

An anonymous referee has provided valuable insight through questions
posed in a report on the first submitted draft of this paper, and for
which the authors are most grateful.\\

TO acknowledges research support from the National Aeronautics and Space Administration (NASA) under Grant No. 80NSSC21K0245 and the National Science Foundation (NSF) under Grant No. AST-2108975. \\

TO is indebted to Andy Baran for suggestions by the pulsation timing technique could be improved. \\

DK thanks the University of the Western Cape for financial assistance.\\

TvH acknowledges research support from the National Science Foundation under Grant No. AST-1715718.\\

MU acknowledges financial support from CONICYT Doctorado Nacional in the form of grant number No: 21190886 and ESO studentship program.\\

This paper uses observations made at the Southeastern Association for Research in Astronomy (SARA-CT) telescope at the Cerro Tololo Interamerican Observatory (CTIO).\\

This paper uses observations made at the South African Astronomical
Observatory (SAAO).\\

This paper includes data collected by the Transiting Exoplanet Survey Satellite (TESS) mission. Funding for the TESS mission is provided by the NASA Explorer Program.\\

Based on observations obtained at the Southern Astrophysical Research (SOAR) telescope under the program allocated by the Chilean Time Allocation Committee (CNTAC), no:CN2020A-87. \\

\vspace{5mm}
\facilities{SAAO: 0.5m, SAAO: 0.75m, SAAO: 1.0m, SAAO:1.9m, SOAR:4.1m, CTIO:0.6 m, TESS}

\software{DoPhot \citep[in an automated form]{Schechter et al.1993}, {Period04 \citep{Lenz2004}}, and {AstroImageJ \citep{Collins et al.2017}}}



\end{document}